# Rotary mechanical circulatory support systems

Milad Hosseinipour[1,2], Rajesh Gupta[3], Mark Bonnell[4] and Mohammad Elahinia[1]

### Abstract
A detailed survey of the current trends and recent advances in rotary mechanical circulatory support systems is presented in this paper. Rather than clinical reports, the focus is on technological aspects of these rehabilitating devices as a reference for engineers and biomedical researchers. Existing trends in flow regimes, flow control, and bearing mechanisms are summarized. System specifications and applications of the most prominent continuous-flow ventricular assistive devices are provided. Based on the flow regime, pumps are categorized as axial flow, centrifugal flow, and mixed flow. Unique characteristics of each system are unveiled through an examination of the structure, bearing mechanism, impeller design, flow rate, and biocompatibility. A discussion on the current limitations is provided to invite more studies and further improvements.

## Introduction

Mechanical circulatory support (MCS) is becoming a realistic alternative to heart transplant for terminal heart failure, after exhaustion of medical and conventional surgical treatments. Each year, there are more patients who need long-term cardiac care. Meanwhile, the number of successful transplants remains stagnant.[1] Several modalities exist for providing support to heart-failure patients as a[2]:

- Bridge to decision (BTD)
- Bridge to recovery (BTR)
- Bridge to transplant (BTT)
- Destination therapy (DT)

Although each device has its unique characteristics, most of the available systems can be classified into two categories: total artificial hearts (TAHs) and ventricular assistive devices (VADs). TAHs replace a portion or all of the heart, similar to a prosthesis, with a device that achieves systole and diastole by filling and emptying artificial chambers in repetitive cycles. VADs, on the other hand, are designed to augment the function of a failing heart as an orthosis without replacing the biological organ. They can provide support to either left (LVAD), right (RVAD), or both ventricles (BiVAD).

Based on the flow regime, VADs can be categorized as: positive displacement (PD-VADs) or continuous flow (CF-VADs). In PD-VADs, the pump is connected to an external driver (pneumatic, hydraulic, or electric) that provides alternating pressure and vacuum to fill and empty the pump chambers. CF-VADs generally provide cardiac support by providing a continuous output flow that increases the arterial blood flow and pressure. Angular momentum is generated through

[1]Dynamic and Smart Systems Laboratory, The University of Toledo, Toledo, OH, USA
[2]Department of Mechanical Engineering, Virginia Polytechnic Institute and State University, Blacksburg, VA, USA
[3]Cardiovascular Medicine Division, The University of Toledo Medical Center, Toledo, OH, USA
[4]Cardiothoracic Surgery Division, The University of Toledo Medical Center, Toledo, OH, USA

**Corresponding author:**
Milad Hosseinipour, Department of Mechanical Engineering, Virginia Polytechnic Institute and State University, Blacksburg, VA, USA.
Email: mhp@vt.edu

rotation of a miniaturized impeller that is further converted to linear flow. The speed of rotation has a direct relation with the output flow rate.

### Continuous flow

While PD-VADs maintain the physiological nature of blood flow, they often have limited lifetime and low reliability due to mechanical failure in diaphragms and valves. Today, CF-VADs comprise over 98% of LVAD implantations in USA,[3] as they often lead to better outcomes.[4] They have shown promising results due to[5]:

- Lower power consumption
- Lower hemolysis risk
- Lower infection risk
- Lower sound level
- Less hospitalization and equipment cost (in long-term)

Moreover, these devices are more compact (25 cc vs. 150 cc) that makes them more suitable for patients with small body-surface area (BSA) and children. The smaller size makes the thoracic implantation achievable, which is a better site than the usual upper abdomen.[5]

LVAD recipients may develop right ventricular failure during or shortly after device implantation. This is partly contributed to rapid unloading of the left ventricle, which may cause deviation of the interventricular septum towards the left ventricle, thus reducing right ventricular efficiency.[6] The very first symptom of right ventricular dysfunction in these patients is often the inability of the LVAD to fill the ventricle that leads to insufficient flow rate. The limited filling is exacerbated when the right heart failure is followed by an elevated transpulmonary pressure gradient or high pulmonary vascular resistance. Until recently, TAHs were the only option for severe biventricular failure. Modern CF-VADs have shown promising results when two devices are implanted as a BiVAD.[7] Patients with BiVAD will not be as dependent on the MCS as patients with TAH. Device malfunctions are easier to manage with a BiVAD compared with a TAH. Moreover, BiVAD implantation does not eliminate the possibility of myocardial recovery.

### Flow regimes

The overall design of the pump including the flow regime, bearings, blood-contacting surfaces, and flow path define the hemolytic and hemodynamic performance of the device. Three types of continuous-flow pumps have been proven in providing durable support:

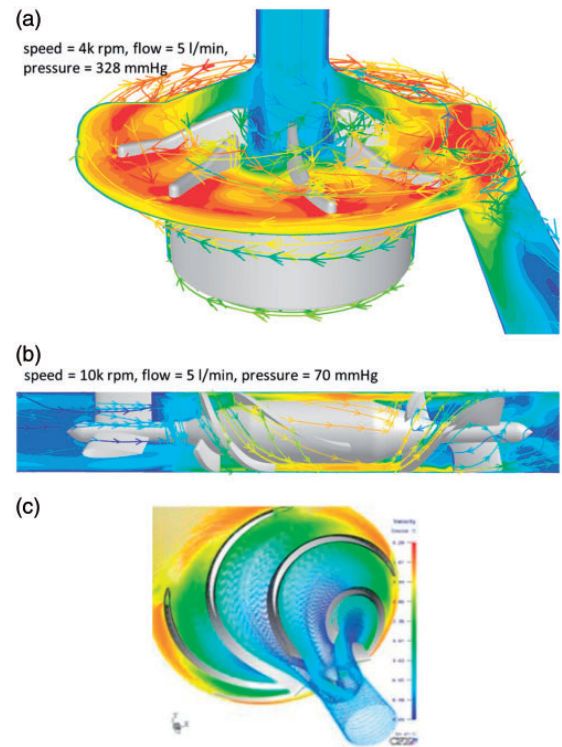

**Figure 1.** Different flow regimes in rotary ventricular assistive devices (reprinted with permission[9,10]). (a) Centrifugal flow, (b) axial flow and (c) mixed flow.

centrifugal flow pump (CFP), axial flow pump (AFP), and mixed flow pump (MFP).[8] In CFPs, the outlet is positioned tangentially to the pump housing, and rotation of the impeller causes the centrifugal force to suck the blood from the inlet and pump it through the outlet. In AFPs, the outlet is co-linear with the rotating section, and the impeller blades are shaped to accelerate the blood both rotationally and axially. An MFP combines both flow types, and impeller blades have a convex profile shape. The degree of convexity varies based on the dominant flow model. Figure 1 provides an example of impeller design and flow regime for each category.

When normalized to the pump size, CFPs are capable of generating a typical pressure gradient, $\Delta P$, over a wide spectrum of flow rates, $Q$. This is often called *flat head curve*.[11] In contrast, AFPs have a steep head curve where there is an almost linear relation between $Q$ and $\Delta P$.[12] MFPs tend to extend the flat region by providing high $\Delta P$ at high $Q$.[13] Figure 2 compares pressure–flow relationships of a typical CFP, AFP, and MFP at the same speed.

The flat head curve means greater sensitivity of the flow–pressure relationship that results in greater change in $Q$ for any given $\Delta P$ across the inlet and outlet of the pump. During a cardiac cycle that the pressure head changes from a min $Q$ at diastole to a max $Q$ in systole (e.g. 40 mmHg to 80 mmHg), the pump exhibits larger change in flow rate (e.g. 0 lpm to 10 lpm), emulating a pulsatile flow. This greater $\Delta Q$ increases the accuracy of the estimated flow from pump speed and power.

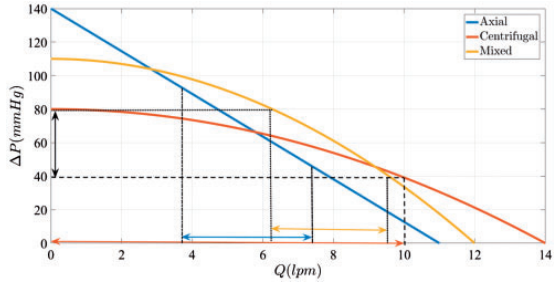

**Figure 2.** Typical P–Q results for centrifugal, axial, and mixed CF-VADs at the same speed.[11,14–16]

## Flow control

In CFPs, the motor current is used as a sensorless index of pump flow and a virtual index of the left ventricular pressures during the cardiac cycle. This is based on the linear current–flow relationship across the full range of operating pump flows in these pumps. This information allows centrifugal pump controllers to monitor the pump flow and the degree of left ventricle unloading by simply monitoring the motor current or power.[11]

The correlation between flow and current in AFPs and MFPs is not nearly linear over the full operating flow range of the pump. Therefore, motor current offers less accuracy for flow estimation. Without a flow sensor, AFPs often cannot report accurate flow, especially at low rates.[12] For example, the HeartMate II axial flow pump does not display flows below 3.0 lpm. The HeartAssist 5 (ReliantHeart) uses an ultrasonic flow probe incorporated in its outlet conduit to measure the flow rate directly.

Regardless of the algorithm, accurate flow estimation depends on viscosity of patient's blood. Currently, only one device (HeartWare HVAD) has adjustable hematocrit setting for adjusting the blood viscosity in flow estimations. Variable viscosity and non-linear relations between flow rate, current, speed, and power, hinder the automatic control algorithms of CF-VADs.

Although CF-VADs do not generate pulsation, the pressure difference between the pump inlet and outlet is pulsatile since the actual heart is still beating. For some devices, backflow may happen at lower speeds (less than 1000 rpm).[17] It has also been reported that the pressure variation can diminish at higher speeds (more than 10,000 rpm) resulting in a non-pulsatile flow.[17] Some researchers have attempted to vary the CF-VAD RPMs throughout the cardiac cycle to achieve some measure of pulsatility.[18] The usefulness of such control algorithms remains to be determined.

Moreover, a CF-VAD is blind to the absolute pressures at the inlet and outlet ports, and responds only to the $\Delta P$. This can lead to inlet cannula suction if the pump increases $\Delta P$ at low left ventricular volume (e.g. hypovolemia). As discussed earlier, AFPs (and MFPs to some extent) have a steep $Q$–$P$ behavior that leads to higher inlet suction at lower flow rate. Higher suction increases the risk of sucking the ventricular wall in around the inlet cannula. This means that a flow control algorithm should include multiple parameters, including (but not limited to):

- Pump speed
- Pressure differential
- Blood viscosity
- Absolute inlet pressure
- Absolute outlet pressure
- Vascular and valvular resistances
- Systemic, ventricular, atrial, and aortic compliances
- Inlet, outlet, and pump resistances
- Aortic inertance
- Inlet, outlet, and pump inertances

The controller state variables are left ventricular pressure, left atrial pressure, arterial pressure, aortic pressure, total flow, and pump flow.

## Bearings

Various bearing mechanisms have been investigated for supporting miniaturized impellers in CF-VADs. Summarized below and illustrated in Figure 3 are those currently in use.

*Mechanical pivot bearing.* In early versions of CF-VADs, the rotor was supported mechanically by journal and thrust bearings.[20] These bearings evolved into low-wear pivot bearings made of precision ceramic components. Figure 3(a) is the HeartMate II (Thoratec) inlet bearing ball and cup. This bearing set was explanted from a pump that was in operation for 4.4 years and demonstrates low wear.[19] Although these bearings are small, they are potential sites for thrombus and fibrin deposition, which is not present in other bearing designs that completely suspend the rotor.[12] The concentration of hydrodynamic loads on these bearings, especially at stress concentration points, makes them theoretically susceptible to wear and fatigue.

*Hydrodynamic radial and thrust bearing.* One of the biggest breakthroughs in the development of MCS systems was the introduction of contactless bearings in 1980s. It was first achieved by filling the rotor bearing with crosslinked blood or other biocompatible materials.[21,22] Hydrodynamic bearing has been utilized in both radial and thrust configurations. In either configuration, a fluid film pressure acts to separate the rotating surfaces from the stationary base. Hydrodynamic bearings provide no life-limiting contact while in motion, however, surfaces will touch when stationary (start and end). Moreover, they apply more shear stress on blood cells especially at the boundary layer. Higher shear stress increases the risk of hemolysis. An example of hydrodynamically suspended pump is HeartAsssit 5 (ReliantHeart). In Figure 3(b), the magnitude and direction of these fluid forces are represented by the height and direction of the arrows.

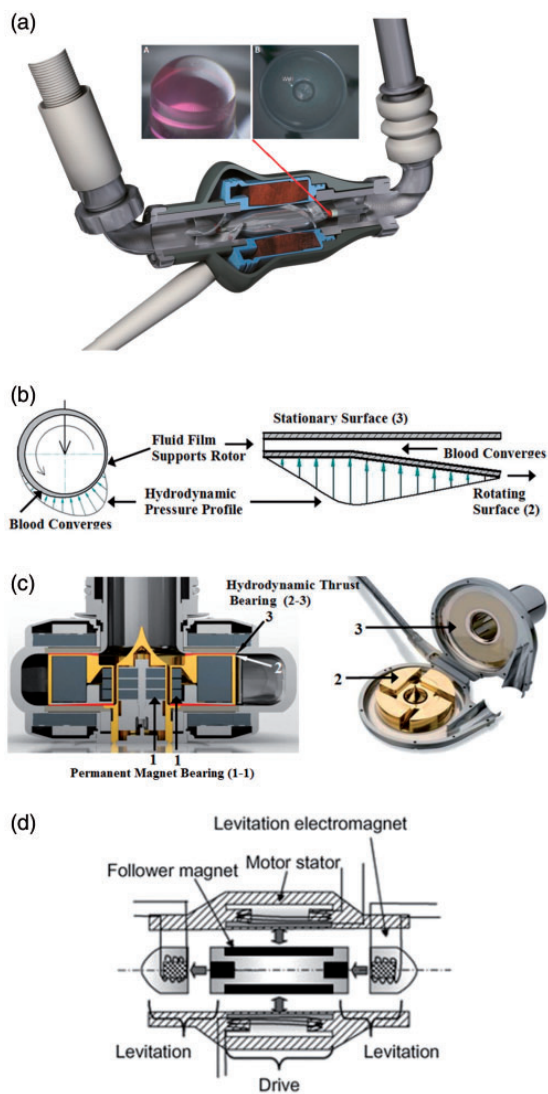

**Figure 3.** Bearing mechanisms in rotary ventricular assistive devices (reprinted with permission[12,19]). (a) Mechanical pivot, (b) hydrodynamic radial (left) and thrust (right), (c) permanent magnet (combined with hydrodynamic bearing) and (d) electromagnetic.

*Permanent magnet bearing.* Using repelling neodymium-boron-iron magnets on the rotor and housing allows for suspending the rotor. Permanent magnet bearings allow for a larger gap between the static motor armature and the rotor, while eliminating the need for mechanical bearings, lubrication, sealing, and purging fluid. However, the magnetic forces are dependent on the instantaneous position of the rotor. Therefore, the magnetic bearings are often used in conjunction with hydrodynamic or electromagnetic bearings. An example of permanent magnet bearing combined with hydrodynamic bearing is HeartWare HVAD (HeartWare International), as illustrated in Figure 3(c). The rotor is aligned and separated from the conical stationary center shaft of the pump housing using magnetic bearing. A thin film of blood generates pressure at the tapered surface of the impeller (red line) to create hydrodynamic elevation.[12] Stator coils in the housing are supplied with electric currents having a frequency and amplitude adjusted to the blood pressure at the pump inlet.[23–25]

*Electromagnetic bearing.* Electromagnetic bearings control three degrees of freedom of the impeller: axial, radial, and tilt. Either axis can be active or passive. In active mode, the repelling force is adjusted by changing the supplied current to the electromagnets based on the instantaneous position or other feedback errors. In passive mode, the supplied current to the electromagnets is constant. The electromagnetic force adjusts automatically based on the position itself. An example of electromagnetic suspension is INCOR (Berlin Heart) (Figure 3(d)). INCOR uses active control for the axial position, and passive control for radial and tilt. Electromagnetic bearings eliminate life-limiting contacts and reduce stress on blood cells at the cost of added components, electronics, and controllers. If a momentary drop in control signal occurs, pump failure will be inevitable. Therefore, these bearings are often backed up with a bearing that does not require active control.

## Biocompatibility

The common practice in manufacturing VADs is highly polished blood-contacting surfaces made of biocompatible materials such as titanium. The sintered titanium surface may be textured to reduce anticoagulation requirements and thromboembolism. To decrease the incidence of thrombus formation, some devices coat the blood-contacting surfaces with materials that reduce platelet activation and adhesion. Common coating materials used in VADs are listed below.

*Carmeda.* BioActive Surface is a heparin-based, thrombo-resistant coating developed for medical devices in contact with blood. The coating is applied by stepwise deposition of cationic and anionic polymers (layer-by-layer) on the device surface, ending with covalent end-point attachment of heparin, a well-established and widely used anticoagulant. What is unique to Carmeda BioActive Surface is the retained functional activity of the immobilized heparin, specifically the capacity to bind to the coagulation inhibitor antithrombin (AT). This natural inhibitor in blood acts by forming a complex with activated coagulation factors (enzymes), thereby neutralizing their pro-coagulant activity. The binding of AT to heparin increases this inhibitory capacity drastically, thus converting AT from a slow to a highly potent coagulation inhibitor. Inactive complexes formed on the immobilized heparin are released and swept away from the surface by the blood flow. Hence, the end-point attached heparin is not consumed but remains active and available for further inhibition. In addition to downregulating the coagulation, the advantages of the Carmeda BioActive Surface include platelet compatibility and reduced activation of natural defense mechanisms.[26]

*Trillium.* It is a polymer coating with applied heparin to the blood-contacting surfaces. A priming layer is bonded to the blood-contacting surface. A hydrophilic functional layer with heparin is deposited to the prime coat and provides the key endothelial-like behavior for the surface of the ventricular assistive device. This functional level includes three primary elements:

- Non-leaching heparin molecules are covalently bonded into the surface to act like heparin sulfate in vascular endothelium.
- Sulfate and sulfonate groups are incorporated into the Trillium biosurface layer to emulate the negative charge of the vascular endothelium. These negatively charged sulphonated polymers repel platelets (that are also negatively charged)[27] and inhibit thrombin by attaching to AT (similar to heparin).[28]
- As a third functional layer, polyethylene oxide polymer is deposited on the surface as a hydrophilic molecule to create an insulating water layer between blood and artificial surface to resist cell adhesion and protein deposition.

These layers lead to a reduction in platelet activation measured by $\beta$-TG. With Trillium, a strict anticoagulation protocol should still be followed and routinely monitored.[29]

To provide a systematic review on implantable CF-VADs, eight prominent technologies are categorized based on their flow regimes. A summary of these devices is given in Table 1. Many other MCS systems have been introduced since 1930s, which are either now defunct or remain investigational. A summary of these devices is provided in Table 2 as a reference. This review is intended to discuss the devices in Table 1 from an engineering point of view, rather than clinical reports. A discussion on the current limitations is provided to invite further developments. Where possible, expert advice has been sought from the manufacturers to provide the most accurate and up-to-date information.

## Axial flow devices

### HeartAssist 5

HeartAssit 5 (ReliantHeart, Inc., Houston, TX, USA) is the modern version of DeBakey/NASA VAD that has been under development since 1988. It consists of an inflow cannula, stator in the pump housing, flow straightener, rotor, and diffuser. This electromagnetically driven pump measures 71 mm long by 30 mm diameter and weighs 92 g. It pumps blood flow of 5 to 6 lpm at an average speed of 10,000 rpm to a maximum of 10 lpm at 12,500 rpm.[30] Due to its small size, HeartAssist 5 may be implanted above the diaphragm, so there is no need to create a pocket for the pump. Much optimization has been done on the inlet angle, outlet angle, axial and radial clearances of blades on flow straightener, rotor (impeller and inducer), and diffuser to minimize the risk of hemolysis and platelet activation. The HeartAssist 5 Pediatric VAD uses a 140° inflow cannula and a 60 mm outflow graft for supporting pediatric patients: as young as four to six years old, BSA larger than $0.7 \, m^2$ and smaller than $1.5 \, m^2$, weight equal or greater than 18 kg. For adults, however, HeartAssist 5 Adult VAD uses an 115° inflow cannula and 90 mm outflow graft. Figure 4 compares the pediatric and adult versions.

A plurality of magnets is placed in each impeller blade that interact with the axially adjusted stator.[31] The straightener and impeller are designed such that they eliminate any area of blood stagnation and prevent thrombus formation. Moreover, there is almost no gap between the body of flow straightener and front hub of the impeller to eliminate flow recirculation zones in contact with stationary parts, which again may be thrombogenic.[21,32] In addition to the pump, the LVAD system comprises a controller module, an ultrasonic flow probe, batteries, a battery charger, and a data acquisition system for home or clinical support. A cable powers the pump percutaneously and sends the feedback to the HeartAssist 5 Conquest Controller. A unique feature of this device is continuous flow rate

Table 1. Summary of commissioned rotary mechanical circulatory support systems.

| Device | Manufacturer | Flow | Side | Location | Suspension and bearings | Size | Output | Approvals |
|---|---|---|---|---|---|---|---|---|
| HeartAssist 5 | ReliantHeart | Axial | LV | Intracorporeal | Hydrodynamic (purged) | 30(D) × 71(L) | 5–10 lpm | FDA (BTT adult) trials, FDA (BTT pediatric), EU (BTT, DT) |
| HeartMate II | Thoratec | Axial | LV | Intracorporeal | Mechanical (blood immersed) | 40(D) × 60(L): 63 mL | 3–10 lpm | FDA (BTT, DT), EU (BTT, DT) |
| INCOR | Berlin Heart | Axial | LV | Intracorporeal | Electromagnetic | 30(D) × 114(L) | 5 lpm | FDA (BTT) trials, EU (BTT, DT) |
| Impella | Abiomed | Axial | LV, RV, BV | Transvalvular | Magnetic and hydrodynamic | 12 Fr to 21 Fr | 2.5–5 lpm | FDA (6h), EU |
| HeartWare HVAD | HeartWare | Centrifugal | LV, RV, BV | Intracorporeal | Electromagnetic and hydrodynamic | 50 mL | 10 lpm | FDA (BTT), FDA (DT) trials, EU (BTT, DT) |
| HeartMate III | Thoratec | Centrifugal | LV | Intracorporeal | Electromagnetic | 69(D) × 30(H): 50 mL | 13 lpm | FDA trials, EU |
| CentriMag | Thoratec | Centrifugal | LV, RV, BV | Extracorporeal | Magnetic | 31 mL | 9.9 lpm | FDA (6 days LVAD and 30 days RVAD), EU (30 days) |
| PediMag | Thoratec | Centrifugal | LV, RV, BV | Extracorporeal | Magnetic | 14 mL | 1.5 lpm | FDA (6 days), EU (30 days) |
| EvaHeart | EvaHeart | Centrifugal | LV | Intracorporeal | Hydrodynamic (purged) | 55(D) × 64(H): 132 mL | 12 lpm | FDA (BTT) trials |
| BPX-80 | Medtronic | Centrifugal | LV | Extracorporeal | Mechanical | 79(D) × 66(H): 86 mL | 10 lpm | FDA (6h), EU |
| BP-50 | Medtronic | Centrifugal | LV | Extracorporeal | Mechanical | 48 mL | 1.5 lpm | FDA (6h), EU |
| Synergy | HeartWare | Mixed | LV, RV, BV | Intracorporeal | Magnetic and hydrodynamic | 12(D) × 49(L): 1.5 mL | 0.3–6 lpm | FDA (BTT, BTR) trials, EU (BTT, BTR) |

BTD: bridge to decision; BTR: bridge to recovery; BTT: bridge to transplant; DT: destination therapy; VAD: ventricular assistive device; TAH: total artificial heart; FDA: Food and Drug Administration; BV: biventricular; LV: left ventricle; RV: right ventricle.

Table 2. Summary of other rotary mechanical circulatory support systems.

| Device | Manufacturer | Flow | Side | Location | Suspension and bearings | Size | Output | Approvals |
|---|---|---|---|---|---|---|---|---|
| Jarvik 2000 | Jarvik Heart | Axial | LV | Intracorporeal | Mechanical (blood immersed) | 25(D) × 55(L) | 3–7 lpm | FDA (BTT, DT) trials, EU (BTT, DT) |
| DuraHeart | Terumo Heart | Centrifugal | LV | Intracorporeal | Magnetic and hydrodynamic | 72(D) × 45(H): 196 mL | 2–10 lpm | FDA (BTT, DT) trials discontinued, EU (BTT, DT) |
| CorAide | Arrow | Centrifugal | LV | Intracorporeal | Magnetic and hydrodynamic | 84 mL | 2–8 lpm | Pre-clinical testing |
| DexAide | Cleveland Clinic | Centrifugal | RV | Intracorporeal | Magnetic and hydrodynamic | 69 mL | 2–6 lpm | Pre-clinical testing |
| TandemHeart | CardiacAssist | Centrifugal | LV, RV, BV | Extracorporeal | Hydrodynamic (purged) | 10 mL | 4 lpm | FDA trials, Canada (Class 4) |
| PediaFlow | PediaFlow Consortium | Mixed | LV, RV, BV | Intracorporeal | Electromagnetic | 28(D) × 51(L) | 0.3–1.5 lpm | In development |
| MiFlow | HeartWare | Mixed | LV | Intracorporeal | Electromagnetic | 22 mL | 2–6 lpm | Clinical trials expected |

BTD: bridge to decision; BTR: bridge to recovery; BTT: bridge to transplant; DT: destination therapy; VAD: ventricular assistive device; TAH: total artificial heart; FDA: Food and Drug Administration; BV: biventricular; LV: left ventricle; RV: right ventricle.

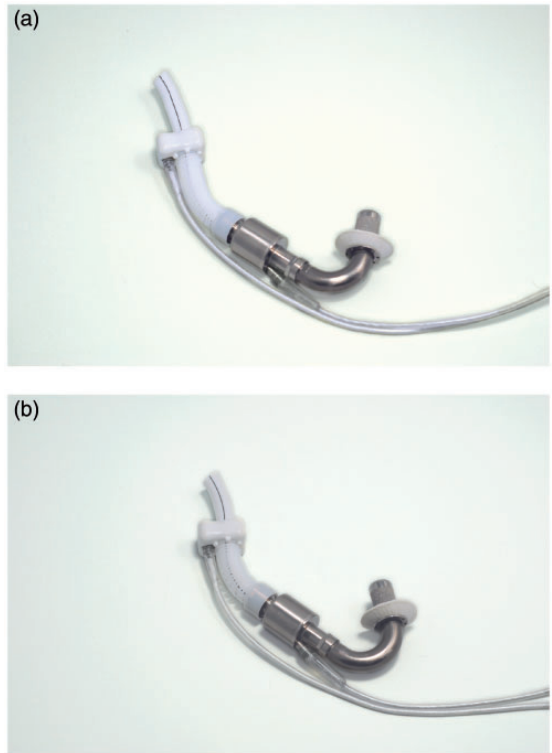

Figure 4. HeartAssist 5 pump (Courtesy of ReliantHeart, Inc.). (a) Adult and (b) low body-surface-area (pediatric).

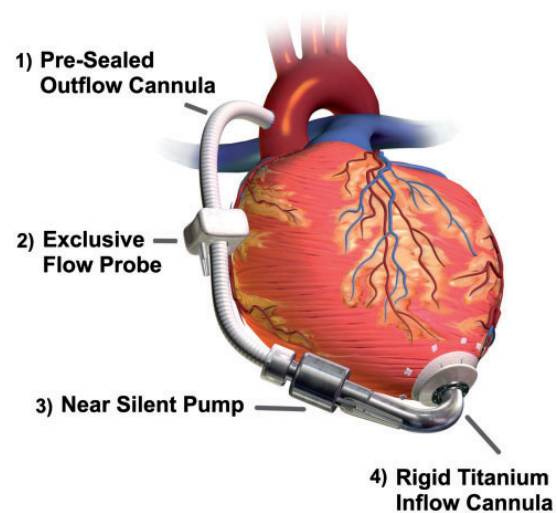

Figure 5. HeartAssist 5 ventricular assist system (Courtesy of ReliantHeart, Inc.).

measurement using an ultrasonic flow probe around the outflow graft (Figure 5).

All blood-contacting surfaces are made of highly polished titanium. In an attempt to decrease the incidence of thrombus formation, all blood-contacting surfaces are coated with Carmeda biocompatible coating. Recently, ReliantHeart designed a remote monitoring system for HeartAssist 5, called HeartAssistRemote, which enables healthcare providers to monitor the pump remotely. So far, HeartAssist 5 is the only remotely monitored VAD in the world. The original DeBakey pump showed a high incidence of thromboembolic events (22%) and pump thrombosis (11–36%) and higher mortality rate in BTT (45%).[33] A recent study showed that using device thrombogenicity emulation (DTE) can optimize the device features. Numerical simulations reported lower stress on platelets and hence lower platelet activation.[34] The pediatric version of HeartAssist 5 received FDA approval for BTT in November 2012, and the Conquest Controller had been given another CE Mark approval earlier in October 2012. In Europe, HeartAssist 5 achieved CE mark approval in 2014. An investigational device exemption (IDE) was granted to HeartAssist 5 in 2014, and clinical trials are in progress.[3]

## HeartMate II

HeartMate II LVAD (Thoratec Corp., Pleasanton, CA, USA and Texas Heart Institute, Houston, TX, USA) is a small AFP for full circulatory support to the left heart failure patients. It is surgically implanted in either pre-peritoneal or intra-abdominal cavities with its 20 mm inflow conduit attached to the ventricular apex and a 14 mm outflow graft connected to the ascending aorta. It provides 3 to 10 lpm continuous flow at a pump speed of 6000 to 15,000 rpm. Sufficient flow rate and hemodynamic pressure are typically achieved at 9000 rpm.[30] The pump housing measures 60 mm long and 40 mm diameter, approximately the size of a D-Cell battery, and weighs 375 g.[35] The rotor, which is the only moving part of the pump, contains a magnet and is powered by an electromagnetic motor to rotate on blood-lubricated bearings.[36] Like the other AFPs, there is a risk of generating negative intraventricular pressure and collapsing the ventricle. Therefore, the position of the inflow cannula and ventricular preload are important. Figure 6 shows HeartMate II with sealed grafts, and Figure 3(a) presents a section cut of the device showing the rotor, magnets, and bearings. All internal blood-contacting surfaces including rotor, inlet stator, outlet stator, and thin-walled duct have a smooth polished titanium surface. The inflow conduit and outflow graft have a textured microsphere surface, which is similar to blood-contacting surfaces on the HeartMate XVE LVAD.

In addition to HeartMate II LVAD implanted inside the body, HeartMate II Left Ventricular Assist System (LVAS) comprises the following components out of body: system monitor, system controller, portable power module, universal battery charger, portable rechargeable batteries, battery clips, and a percutaneous driveline.[37] The polyester velour covered driveline exits the skin and connects the motor to the system controller, which continuously powers and controls the whole system. Figure 7 shows an X-ray of a patient with HeartMate II LVAS. The system may be driven by the portable batteries or the power module independently. The controller runs the pump at fixed-speed mode, and the speed of rotation is determined at the time of implantation based on hemodynamic needs. The rechargeable batteries are 14 V Li-Ion or 12 V NiMH, and each pair can provide enough energy to let the patient live tether free for up to 10 h. The 20-foot power cable in the power module makes it possible to run the system using external AC power source.

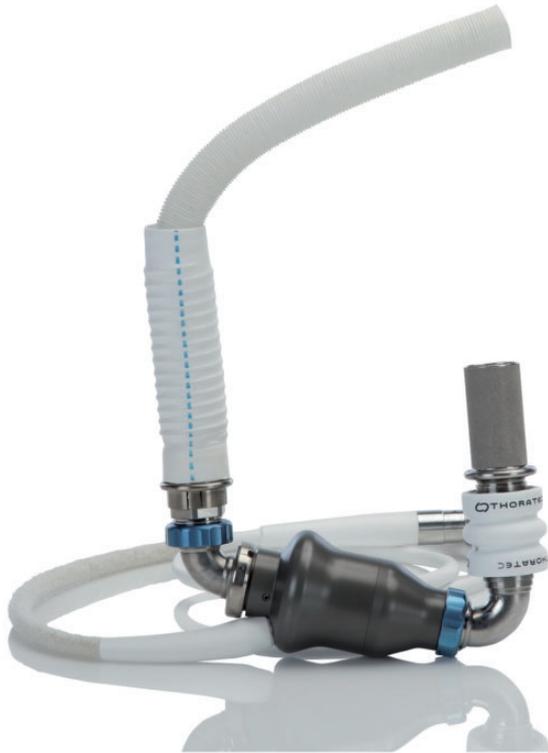

**Figure 6.** HeartMate II pump with sealed grafts (Courtesy of Thoratec Corp.).

The portable power module runs the pump and powers the controller. It has integrated battery for emergency support for up to 30 min. The system monitor communicates with the system controller and allows the user to monitor system parameters or change the operation speed. It also provides an estimation of output blood flow based on the pump speed and amount of power being given to the pump. The relationship between power and flow at any speed is mostly linear, with the exception of low and high-end regions. Currently, Thoratec (Pleasanton, CA, USA) and WiTricity (Watertown, MA, USA) engineers are developing a fully implantable version of the device using an optimized transcutaneous energy transmission system (TETS).[35]

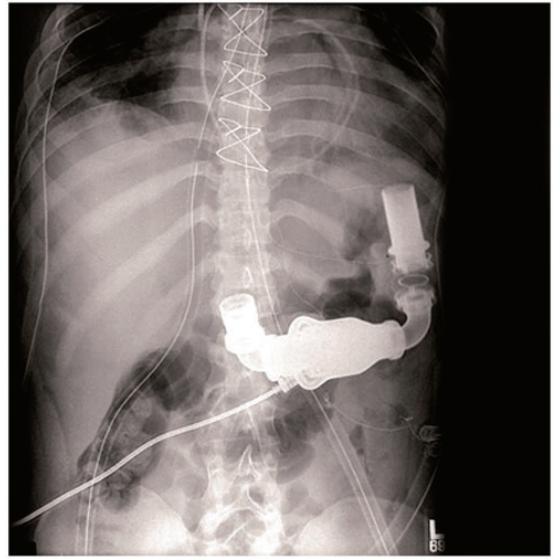

**Figure 7.** X-Ray of a patient with HeartMate II LVAS (Courtesy of Thoratec Corp.).

HeartMate II has been approved in USA and Europe for BTT (2008) and DT (2010). In 2009, a seminal trial showed better outcomes for HeartMate II LVAD compared with HeartMate XVE in a cohort of patients as ill as those in the original REMATCH trial. HeartMate II LVAD is the most widely used LVAD in the world with more than 12,969 implantations between 2004 and 2012.[3] In 2015, Thoratec, Inc. announced the 20,000th implantation.

### INCOR

INCOR (Berlin Heart GmbH, Berlin, Germany) is a long-term implantable axial continuous-flow ventricular assistive device with direct drive and magnetically levitated rotor. Complete suspension of the rotor with two electromagnetic bearings at each end eliminates the need for mechanical bearings, lubrication, sealing, and purging fluid. INCOR measures 30 mm outer diameter and 114 mm long, somehow the longest among axial flow devices.[30] It weighs 200 g mostly contributed to its magnetic system components. The pump is capable of providing a maximum flow rate of 5 lpm at rotation speed between 5000 and 10,000 rpm. In such range, the pump consumes 2–4 W power against 100 mmHg pressure. Silicon inflow cannula delivers blood from left ventricular apex, and another silicon outflow cannula is anastomosed onto the ascending aorta. All blood-contacting surfaces are heparin coated by Carmeda process.[38] A percutaneous cable connects the pump to control unit and battery packs out of the body. Figure 8 shows the latest version of INCOR.

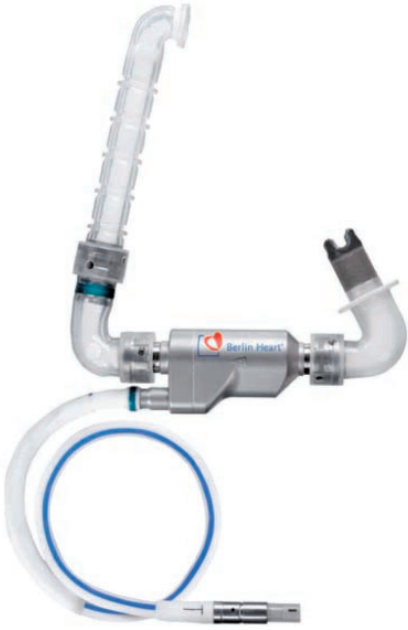

Figure 8. INCOR pump (Courtesy of Berlin Heart GmbH).

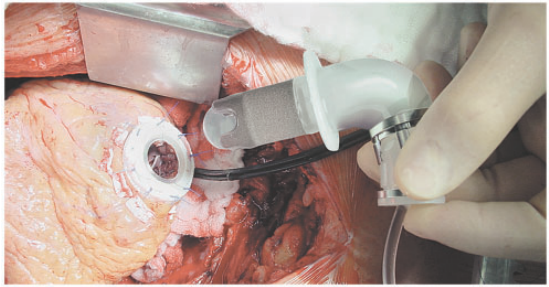

Figure 9. Insertion of inlet cannula of INCOR into the left ventricular apex (Courtesy of Berlin Heart GmbH).

All components of the INCOR system are interconnected with snap-in connectors for easier assembly. Figure 9 was taken during the insertion of INCOR inlet cannula into the apex. The pump housing has a stationary flow straightener at inflow port and a stationary diffuser at outflow port, each equipped with magnets and magnetic coils. The flow straightener and diffuser reduce the spin of blood flow and add to the pressure gradient across the pump. The impeller has arrangements of magnets inside it and is enclosed by the motor stator. Radial and tilting impeller movements are passively controlled. A unique controller algorithm positions the rotor axially within the titanium housing to create a constant laminar flow. This controller commands the currents required for impeller rotation and stabilization. The faster the pump rotates, the higher pressure gradient it generates across the pump, leading to a greater tendency of the impeller displacing from its central position. The required current to counterbalance this flow force is the principle variable of the pump flow controller.[39]

The first human clinical trial of INCOR started in 2002. Out of 212 cases worldwide, 65 (31%) were bridged to transplantation, 11 (5%) were bridged to recovery, and 43 (20%) were still on the device at the time of publishing the results.[40] Blood chemistry found no trauma or hemolysis, and there were no device-related mortalities. INCOR has been approved to be used for BTT and DT in Europe since 2013. Berlin Heart started clinical trials for INCOR in 2009 under an FDA IDE approval.

### Impella

Impella (Abiomed, Inc., Danvers, MA, USA) is one of the world's smallest pumps and technically an expansion of Hemopump (Medtronic, Inc., Minneapolis, MN) that works based on the principle of Archimedes screw.[41] This miniature device, often called transvalvular pump, is small enough to be implanted into the ventricle through the femoral artery and powered via a flexible cable from an external (percutaneous) source of energy. Impella is developed to serve the patients who develop cardiogenic shock, and standard medical therapy has not been able to control or recover the complication. It is designed to restore stable hemodynamics, reduce infarct size, and protect the myocardium from ischemic damage. An impeller is levitated magnetically without any mechanical bearing. It is passively centered in the pump casing via permanent magnets in combination with hydrodynamically acting driving forces. The lateral centering of the impeller is also affected by the permanent magnets cooperating with the external driving magnetic means. Rotation of the impeller in the outlet area pulls blood from the inlet cannula into the root of the ascending aorta. The largest diameter section of the device is motor housing that resides in the aortic root, integrated with a cannula portion.[42] The small diameter cannula passes through the aortic valve with the valve leaflets around it. It should be small enough to minimize the risk of aortic valve regurgitation. Impella has four versions: Impella 2.5, Impella 5.0, Impella LD (left direct), and Impella CP (cardiac power). A brief comparison of the three versions is given in Table 3.

Impella 2.5 is implanted percutaneously via a 13 Fr sheath in the femoral artery, then ascending aorta, across the valve (aortic or tricuspid) and then into the ventricle.[43] Impella CP delivers more flow on the same platform as Impella 2.5. Impella 5 is inserted peripherally into the femoral or axillary artery into the left ventricle. Impella LD, however, is inserted surgically directly into the ascending aorta. Figure 10 shows the Impella 2.5 and an illustration of the implanted device. All devices are controlled with Impella Console (Abiomed, Inc., Danvers, MA, USA). On each version, a pressure sensor is used to monitor catheter position and calculate the flow.[44] Purge tubing attached to the pump supplies the pump bearings with a blood-compatible purge fluid, typically 20% dextrose in water. The increased viscosity of this cleansing fluid helps to create a pressure barrier against the blood that the device is exposed to, thereby minimizing thrombus formation.[45] A pressure transducer (CM-Set) measures the purge pressure in the purge lumen of the catheter. Currently, B. Braun (Melsungen, Germany) purge pump is a separate unit in the Impella Console System.

Table 3. Differences between Impella 2.5, Impella 5.0, Impella LD (left direct), and Impella CP (cardiac power).

| Version | Impella 2.5 | Impella 5 | Impella LD | Impella CP |
| --- | --- | --- | --- | --- |
| Catheter | 9 Fr | 9 Fr | 9 Fr | 9 Fr |
| Motor size | 12 Fr | 21 Fr | 21 Fr | 14 Fr |
| Flow rate | 2.5 lpm | 5 lpm | 5 lpm | 4 lpm |
| Support | Five days | Seven days | Seven days | Five days |
| Max speed | 51,000 rpm | 33,000 rpm | 33,000 rpm | 51,000 rpm |
| Guide wire | 0.046 × 260 cm | 0.064 × 260 cm | Not Required | 0.046 × 260 cm |
| Pressure sensor | Direct | Differential | Differential | Direct |

FDA granted 510(k) clearance to Impella 2.5 (2008), Impella 5 (2009), Impella LD (2009), and Impella CP (2012) for partial circulatory support up to 6 h in cardiac procedures not requiring cardiopulmonary bypass. In 2016, Abiomed, Inc. announced that it has received FDA Pre-Market Approval (PMA) for the Imeplla product series. Altogether, Impella family has been utilized to support more than 10,000 patients in USA. In Europe, Impella family has received CE mark and is widely used.[42]

### Centrifugal flow devices

#### HeartWare HVAD

HeartWare HVAD (HeartWare International, Inc., Framingham, MA, USA) is a miniaturized centrifugal flow pump for full circulatory support. Figure 11 approximates the size of HeartWare HVAD. Although mainly designed for left heart, it has also been modified for biventricular support.[46,47] The impeller is the only moving part of the pump and is suspended using a hybrid suspension mechanism. The hybrid suspension incorporates passive magnets and hydrodynamic thrust bearings. A gentle inclination is created on the upper surface of the impeller blades. As the impeller rotates, the generated blood flow across these inclinations creates a permanent gap between the impeller and housing such that there is no mechanical contact between the impeller and pump housing. This hybrid suspension provides an effective wearless operation, which significantly downsizes the pump. Figure 3(c) shows an internal view of HeartWare HVAD.

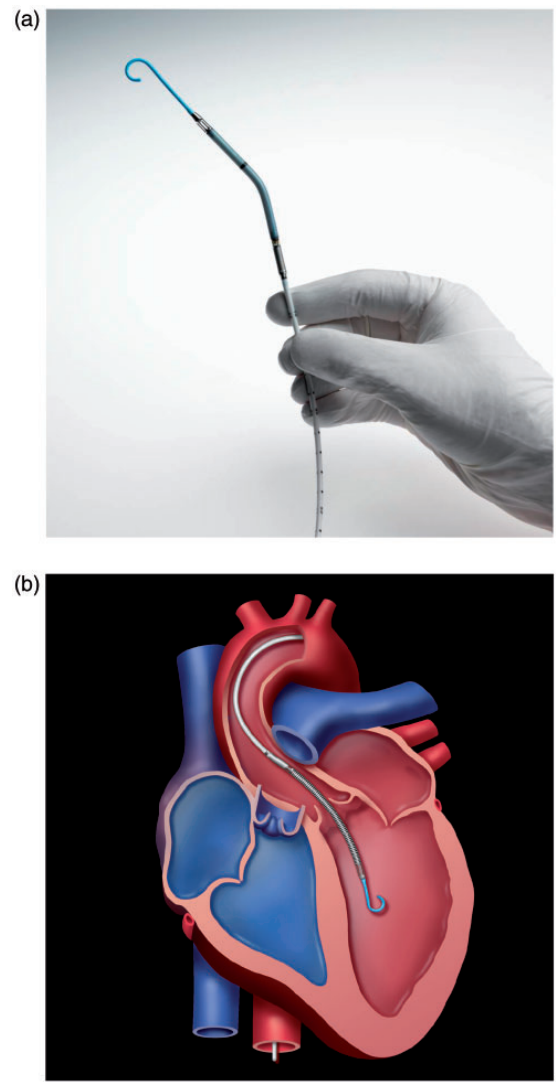

**Figure 10.** Impella LVAD (Courtesy of Abiomed, Inc.). (a) Impella 2.5 and (b) Illustration of the implanted device (pediatric).

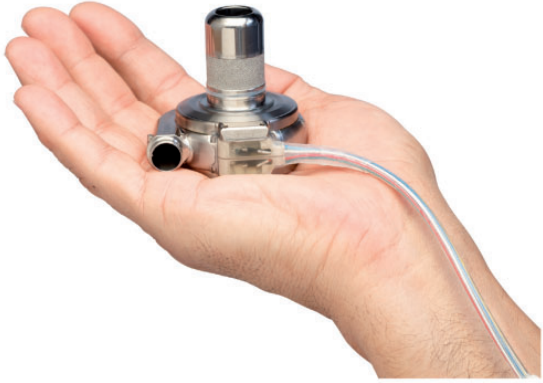

Figure 11. Approximate size of the HeartWare HVAD (Courtesy of HeartWare International, Inc.).

HeartWare HVAD weighs 145 g and measures 50 mm diameter and 50 cc displacement volume. The impeller within the titanium housing rotates in the range of 2400 to 3200 rpm and is capable of delivering up to 10 lpm flow rate. Unlike similar devices, it is designed to be implanted above the diaphragm adjacent to the heart, thereby eliminating the need for abdominal surgery. The inflow of the pump is taken via an integrated inflow cannula, which is directly inserted into the ventricular apex. The pump then propels blood through an outflow graft, equipped with strain relief, into patient's ascending aorta. Figure 12 illustrates the implantation of HeartWare HVAD. Two independent motor stators with separate circuitry are available in the pumping unit to easily switch between single and dual stator modes and therefore increase the device reliability.

A single driveline exits the patient's body and connects the pump to an external controller and battery packs. Either two rechargeable batteries or a battery plus an adapter (AC or DC) may run the controller and pump. In the latter case, the pump may be plugged into an AC or DC outlet. One of the batteries runs the system for 6 h. When depleted (or adapter disconnected), the controller switches to the backup battery and the depleted one should be recharged. In 2012, HeartWare, Inc., announced its partnership with DUALIS MedTech GmbH (Seefeld, Germany) to develop a fully implantable version of HVAD using transcutaneous energy transmission systems. The manufacturer is also developing a miniaturized AFP, with an approximate displacement volume of 15 cc, called HeartWare MVAD.[48] MVAD impeller suspension is based on the same technology used in HVAD and rotates in the range of 16,000 to 28,000 rpm within a titanium cylindrical housing, capable of providing up to 10 lpm flow rate. Ideally, MVAD will be implanted without the need for median sternotomy. The outflow of MVAD is a 10-mm double woven graft anastomosed to the aorta.

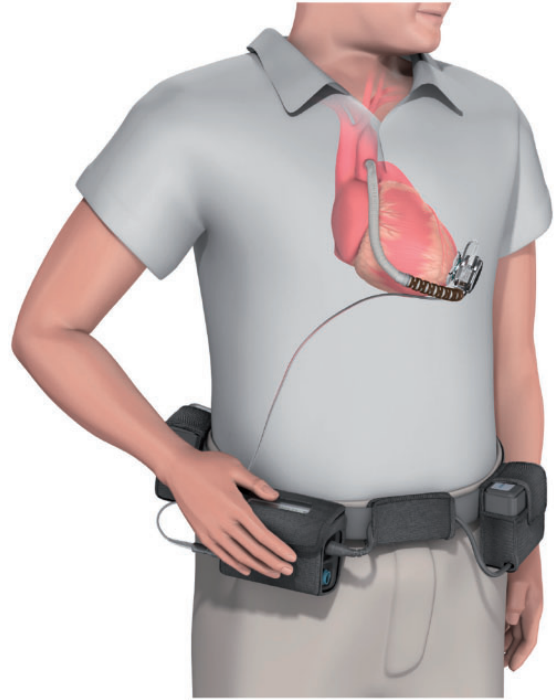

Figure 12. Illustration of the HeartWare HVAD system (Courtesy of HeartWare International, Inc.).

In 2010, the results of the company's ADVANCE trial led to approval of HeartWare HVAD as a BTT.[49] Results showed survival to transplant or ongoing support of 92% after 180 days. HVAD is more compact than HeartMate II, which enables its implantation in smaller patients. This pump is designed for intra-pericardial placement that decreases the risks of infection or other complications that may be associated with intra-abdominal or pre-peritoneal approaches.[7] HeartWare, Inc. is currently conducting a clinical trial on HeartWare HVAD for DT.

### HeartMate III

HeartMate III (Thoratec, Pleasanton, CA) is a continuous flow centrifugal pump with a magnetically levitated rotor. The pump is designed to be able to mimic the biological pulsatile flow through rapid changes of the pump speed. Although not unique to HeartMate III, this is a highly important feature considering that the long-term effects of continuous flow devices on organ perfusion are not well understood yet. HeartMate III works based on a self-bearing or bearingless electromagnetic design to minimize the life-limiting friction and wear. A recent design of the pump measured 69 mm outer diameter by 30 mm height and 200 g weight, which includes the motor, inflow cannula, flexible recovery section, outflow graft, bend relief, and all connecting hardware (Figure 13). Rotating in the speed range of 2000 to 5500 rpm with the prime volume of 50 cc, the pump is capable of providing up to 13 lpm flow rate. The major parts of the pump are[50]:

- An upper housing that includes the inflow/outflow cannula and the top half of the volume
- A lower housing that includes the bottom half of the volume and a cavity for the motor
- A motor that includes drive and levitation functions in a single magnetic structure
- A rotor that includes an impeller and passive permanent magnets

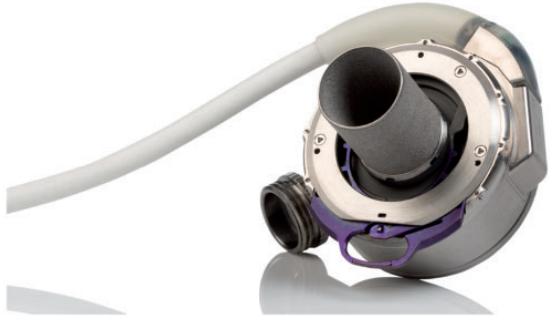

Figure 13. HeartMate III (Courtesy of Thoratec Corp.).

The spinning rotor draws blood into the pumping chamber along the rotor's axis and propels it tangentially through an outflow graft aligned perpendicular to the inflow cannula.[51] Motor drive and magnetic levitation coils share the same stator. Thus, no separate motor or bearing exists. The impeller rotation is caused by a moving magnetic field generated by the drive coils.[52] The rotation and radial levitation of the rotor are controlled by a two-axis active control system and a feedback control loop. The remaining degrees of freedom (axial and tilting) are controlled passively by the magnetic support. Inflow cannula of the device is inserted into the left ventricular apex, and the outflow graft is anastomosed to the ascending aorta. All blood-contacting surfaces of the pump are made of titanium with the exception of polytetrafluoroethylene (PTFE) washers at joints and woven polyester grafts. Like other members of the HeartMate family, HeartMate III incorporates sintered titanium-textured surfaces to reduce anticoagulation requirements and thromboembolism. Both percutaneous and implantable versions of the device are under development. For the percutaneous version, a single driveline exits skin and connects it to a belt-mounted driver/controller. Using a pair of rechargeable batteries, the wearable power transmission can run the system for up to 6 h.[53] The design of the HeartMate III has been modified several times during the developmental process. In another modified version of the pump, the inflow cannula is integrated into the pump housing and is inserted directly into the left ventricular apex.[52] Using full active control over the rotor ramp speed, it can also provide an optional near-physiologic artificial pulse. The Momentum 3 US IDE clinical trial for HeartMate III is ongoing with more than 1000 participants to compare the performance of HeartMate III to that of HeartMate II. In 2015, Thoratec announced that HeartMate III has met the primary endpoint of the CE Mark trial. This trial was a single arm, prospective, multicenter, non-blinded and non-randomized study. The study involved up to 50 patients at nine sites in Europe, Australia, and Canada.[54] DT and BTT trials for Heartmate III are currently ongoing.

## CentriMag and PediMag

Thoratec CentriMag and PediMag (Thoratec Corp., Pleasanton, CA, USA) are a family of extracorporeal quickly installable devices for short-term left/right or biventricular use in patients who need cardiopulmonary support with potentially recoverable heart failure. CentriMag and PediMag pumps use magnetic levitation to suspend the rotor by eight L-shaped iron cores.[41] Their extracorporeal system comprises the centrifugal pump, an electromagnetic motor, a drive console, an ultrasonic flow probe, and a tubing circuit.[55] To reduce hemolysis, mechanical gaps in the pump are more than 0.6 mm and the pump has no valves, seals, or moving parts other than the rotor.[56] CentriMag is appropriate for patients larger than 20 kg and is capable of providing up to 9.9 lpm flow rate under normal physiologic pressure at rotor speed up to 5000 rpm with a priming volume of 31 mL. PediMag is optimized for pediatric patients around 20 kg and is capable of 1.5 lpm flow rate with a priming volume of 14 mL. Figure 14 shows the Thoratec PediMag (a) and CentriMag (b). Figure 15 shows the Thoratec CentriMag installed on the electromagnetic motor.

Both CentriMag and PediMag operate on the same hardware platform, eliminating the need to have separate system setups to support adult and pediatric patients. They can be used as right/left or biventricular assist devices with cannulation of the left ventricular apex, left atrium, or right atrium for inflow, and aorta or pulmonary artery for outflow.[55] In both pumps, the inlet is on the same axis as the rotor, while the outlet is directed perpendicular to the inlet, tangent to the pump circle. The inflow is a 32 Fr wire-reinforced cannula, whereas the outflow cannula is a 22 Fr straight one.[57,58] These pumps are currently used for stabilizing hemodynamic needs with minimum mechanical failure and blood-related problems such as hemolysis or pump-induced trauma. Figure 16 compares the implantation of CentriMag with HeartMate II.

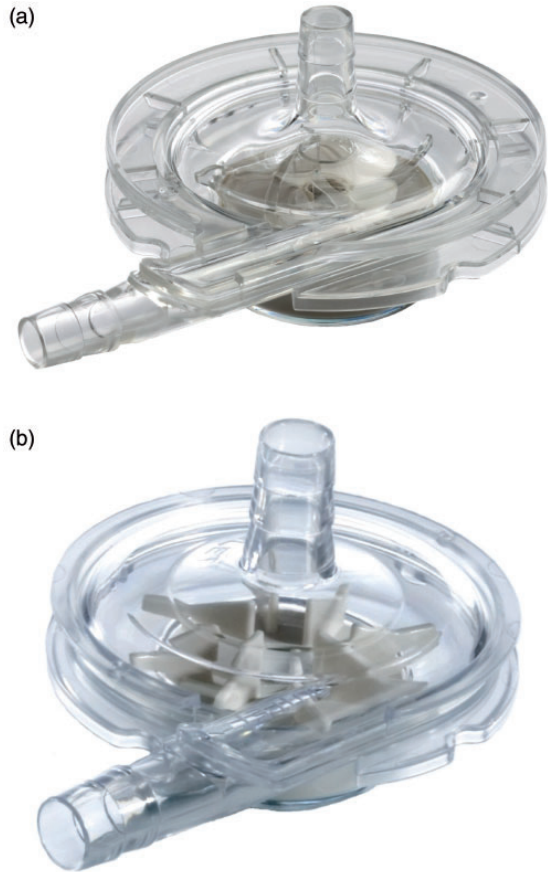

**Figure 14.** Thoratec PediMag and CentriMag BiVADs (Courtesy of Thoratec Corp.). (a) PediMag and (b) CentriMag.

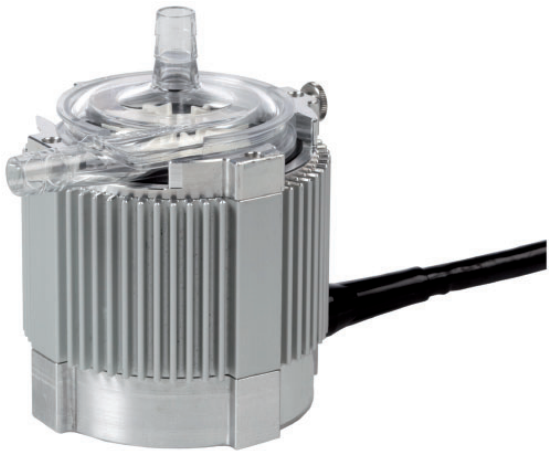

**Figure 15.** Thoratec CentriMag installed on the electromagnetic motor (Courtesy of Thoratec Corp.).

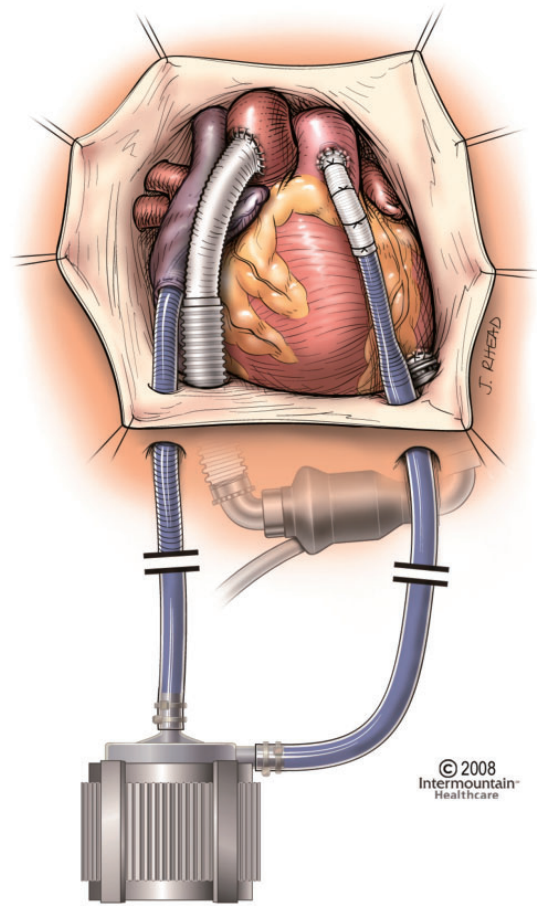

**Figure 16.** Implantation of Thoratec CentriMag versus HeartMate II (Courtesy of Thoratec Corp.)

FDA granted 510(k) clearance to CentriMag for use up to 6 h in patients requiring extracorporeal circulatory support during cardiac surgery. Investigational trials are ongoing in an US pivotal trial to demonstrate safety and effectiveness for periods of support of up to 30 days. Additionally, CentriMag is approved under an FDA Humanitarian Device Exemption to be used as an RVAD for periods of support up to 30 days in patients with cardiogenic shock due to acute right ventricular failure. In Europe, CentriMag has CE Mark approval for up to 30 days of support.

Both devices are cleared for clinical use for up to 6 h. CentriMag is also being evaluated for BRT or BTT therapy. It is approved for RVAD support for up to 30 days for patients with cardiogenic shock due to right ventricular failure. In conjunction with CentriMag console and motor, PediMag blood pump is 510(k) cleared by FDA for support periods of up to 6 h. Thoratec announced that they have submitted an IDE to the FDA to begin a U.S. clinical trial examining the safety and probable benefit of the device for use up to 30 days to support pediatric patients. Outside USA, the device is branded as PediVAS and has CE Mark approval for support durations of up to 30 days.

### BPX-80 Bio-Pump Plus and BP-50 Bio-Pump

BPX-80 Bio-Pump (Medtronic, Minneapolis, MN, USA) is a continuous extracorporeal CFP made of molded polycarbonate for short-term supports. It has no impeller vane or roller, which is not commonly found in other centrifugal pumps.[59] The constrained-forced vortex pumping principle is based on a series of smooth-surfaced rotating cones that pull the blood into the vortex created by the rotation of cones. As the blood flows toward the outlet, the vortex energy, generated by these cones, is transferred to the blood in the form of pressure and velocity. The smooth vortex cone design of BPX-80 Bio-Pump has shown greater air retention compared with impeller-based centrifugal pumps. Higher air retention means less air passes through the pump outlet and finally towards the patient. This pumping principle promotes laminar flow and improves the blood-handling capabilities. The pump utilizes a polycarbonate journal bearing to support the impeller, which may lead to lower durability and higher mechanical wear. The pump also comprises a double-lip seal design over this precision bearing. The size of the pump is as follows: cone-shaped impeller is 79 mm diameter, top housing is 50 mm high, and the bottom housing is 16 mm thick. BPX-80 Bio-Pump has a priming volume of 86 mL, weight of 200 g, and consumes 8 W power.[13] It is capable of providing up to 10 lpm flow rate at rotation speeds of up to 4500 rpm.[59,60] Figure 17 shows the BPX-80 Bio-Pump Plus and its cutaway section.

To decrease flow recirculation at the outlet opening and reduce the free plasma hemoglobin, the ellipse-shaped intersection of the outlet connector and pump housing is straightened to a nearly vertical line, based on computational fluid dynamics (CFD) analysis of shear stress near the opening. In addition to that, the pump outlet is tapered in order to create a more gentle blood flow. The pump is available with two biocompatible coatings: Carmeda BioActive Surface (under license from Carmeda AB, Sweden.) and Trillium Biosurface (under an agreement with BioInteractions, Limited, United Kingdom). The BPX-80 Bio-Pump Plus runs in conjunction with the remote-tethered Bio-Console 560 drive system that runs the pump using integrated brushless DC motor, powers the pump by two 12 V lead-acid gel rechargeable batteries, and controls the pump flow rate. It also detects generated bubbles and measures the blood pressure. The Bio-Console560 weighs 17.19 kg and measures $31.88 \times 22.38 \times 43.02$ cm, providing a limited level of mobility. Figure 18 shows the Bio-Pump BP-80 and Bio-Console560.

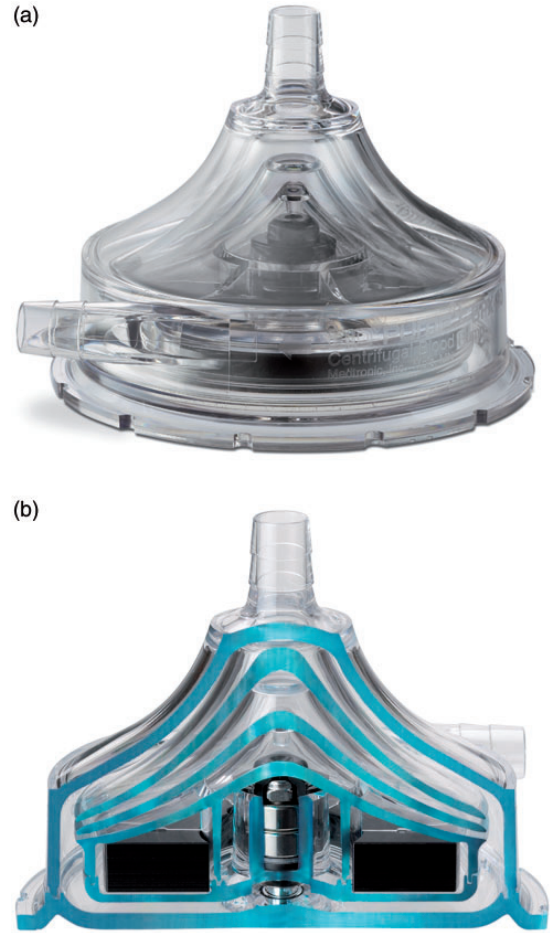

**Figure 17.** BPX-80 LVAD (Courtesy of Medtronic). (a) BPX-80 Bio-Pump Plus and (b) cross section.

BP-50 Bio-Pump (Medtronic, Minneapolis, MN, USA) is similar to BPX-80 Bio-Pump Plus, designed for pediatric use and patients with special needs. With a 48 mL priming volume, it is capable of providing up to 1.5 lpm flow rate with a maximum speed of 4500 rpm. It may be controlled by the same controlling units that run BP-80 Bio-Pump Plus such as Bio-Console 560. Similar to BP-80, BP-50 is available with Carmeda Bioactive Surface. Figure 19 shows the BP-50 Bio-Pump and its particular console called Bio-Console 540.

BPX-80 Bio-Pump Plus is a 510(k) cleared device since 1985. It is approved to be used in conjunction with Bio-Console up to 6 h in extracorporeal cardiopulmonary bypass. BP-50 has also got a 510(k) clearance

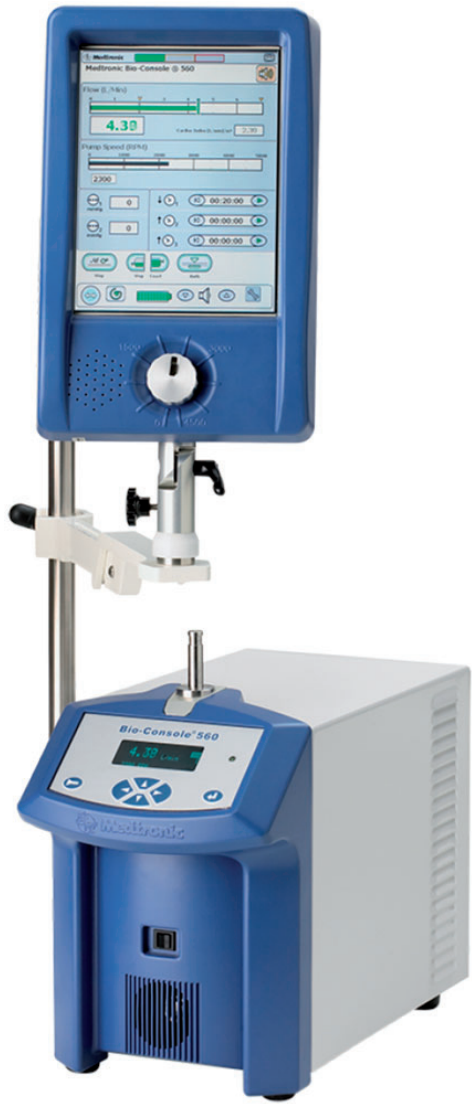

Figure 18. Bio-Pump BP-80 and Bio-Console560 (Courtesy of Medtronic).

from FDA. BPX-80, BP-50, and BP-Console are all CE marked in the Europe.

### EvaHeart

EvaHeart (Evaheart, Inc., USA, Pittsburgh, PA, USA developed by Sun Medical Technology Corp., Nagano, Japan) is an implantable CFP designed for long-term left ventricular support. It incorporates one water-lubricated hydrodynamic journal bearing with pure, sterile water injected into the pump housing from an external water reservoir in the controller package via two 2.5 mm pipes embedded in the percutaneous driveline. The water and blood chambers are separated from each other by a thin seal called Cool-Seal. The water lubricates the journal bearing, cools the motor coil and journal bearing down, and flushes the inner seal faces. The hydrodynamic journal bearing supports a shaft that connects the motor to the impeller. The pump measures 55 mm by 64 mm with a priming volume of 132 mL and weighs 420 g. It is capable of providing up to 12 lpm flow rate at 2600 rpm against 100 mmHg pressure. At flow rates within 5 to 9 lpm, the power consumption of the pump is reported 9 to 10 W.[61] To provide complete bearing wash out during operation, vanes have an open-faced and swept back shape.[2] Sixteen millimeters inflow cannula and outflow graft are connected to the ventricular apex and ascending aorta, respectively. They are made of ePTFE or polyester, and all blood-contacting surfaces are coated with either diamond-like carbon or a proprietary polymer called MPC.[62] Although the pump has a completely flat pressure–flow curve, the pump output may become pulsatile. During systole, when the pressure difference between left ventricle and aorta drops, an instant high peak flow leads to a higher peak pressure in the aorta, which is identical to systolic pressure. During diastole, as the left ventricle-aorta pressure difference increases, pump flow rate decreases and thus creates an instant lowest pressure.[63] The initial clinical trial in Japan completed in 2008 after successful implantation in 18 patients. The mean support duration was 2.76 years with a maximum of 6.29 years. Eight patients were successfully bridged to transplantation with mean support time of three years. The overall survival at six months was 89%.[64] The device received the final regulatory approval from the Japanese Pharmaceuticals and Medical Devices Agency. As of December 2013, 118 patients have received the device (80 ongoing). EvaHeart is still limited to investigational purposes by Federal law in USA. Currently, Evaheart, Inc. is conducting BTT trials under an FDA-approved IDE.[3]

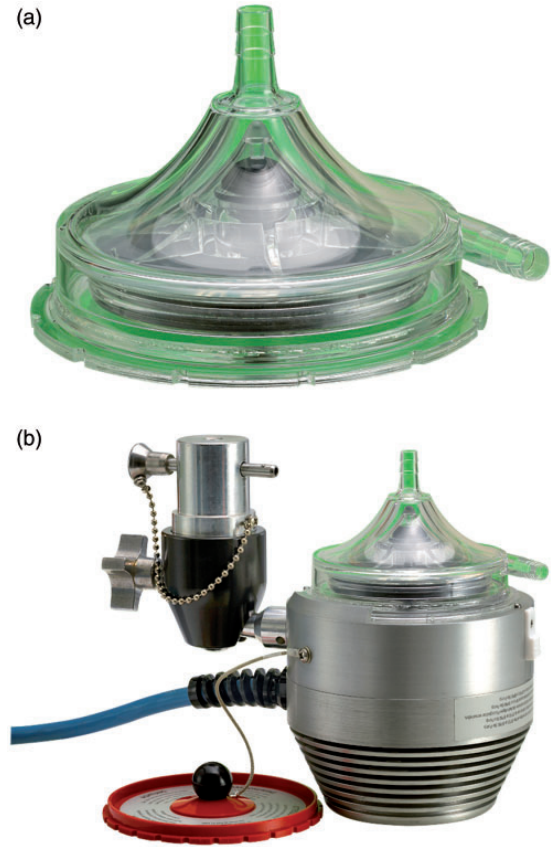

Figure 19. BP-50 LVAD (Courtesy of Medtronic). (a) BP-50 Bio-Pump and (b) Bio-Console 540.

## Mixed flow devices

### Synergy micro-pump

Synergy Micro-Pump (CircuLite, Inc., Saddle Brook, NJ, USA) is a miniaturized partial circulatory support system developed for long-term use in NYHA class IIIb/early IV patients. It is intended to be implanted in less ill patients who have symptomatic heart failure, but BTR is preferred over BTT or DT for them. Synergy will augment the native ventricle that provides some flow by itself, but not sufficiently. This allows the heart to rest and recover, until the appropriate time for device explantation.[2] Synergy is often characterized by its hybrid or mixed flow that presents both features of AFPs and CFPs. It combines axial, centrifugal, and orthogonal flow paths with a single stage impeller that is driven by an integrated brushless DC motor. This titanium micro-pump measures 12 mm diameter by 49 mm length, the approximate size of an AA battery. The pump weighs 25 g and has a priming volume of 1.5 mL. Figure 20 approximates the size of Synergy. This small size allows the device to be implanted surgically in the subclavicular space, into a pacemaker-like pocket, through a right upper chest thoracotomy. This eliminates the need for cardiopulmonary bypass or sternotomy and makes the implantation procedure less invasive. The inflow cannula is made of silicon and reinforced by NiTi shape memory alloy and is inserted into the left atrium. The outflow graft is made of PTFE with an inner diameter of 8 mm and thickness of 1 mm, anastomosed to the subclavian artery. A small percutaneous lead connects the pump to the external wearable controller that weighs around 1 kg. The rechargeable dual battery pack can run the system for 6 to 8 h. The pump's rotor is supported by a mechanical pivot bearing and stabilized by a combination of magnetic and hydrodynamic forces. By operating at the range of 20,000 to 28,000 rpm, the pump is capable of providing 2 to 4.25 lpm flow rate. Although the small size and achieved an operational speed of the device are promising, this lower flow is a concern in increasing the risk of thrombus formation.[65] Synergy received CE Mark in 2012 and an IDE approval in 2013 for investigational trials. The device is being supported by an NIH grant for further developments, system validation, and clinical evaluation. The manufacturer is planning to expand the Synergy technology to cover a broader patient population. Thus, the next-generation of the Synergy pumps is currently under development. These modified versions are designed for endovascular implantation, right heart support, all support, child support, and infant support. Figure 21 compares the implantation of these pumps. The endovascular system will be implanted with the inflow cannula placed trans-septal into the left atrium from the subclavian vein and the outflow graft anastomosed to the subclavian artery. Preclinical animal studies of this device are now completed, and it is approaching clinical evaluation in Europe. A modified Synergy pump is also being designed for right heart support or working in conjunction with the current design as a biventricular partial circulatory support system.[66] Animal studies and cannula development of this device are in progress. All-support platform is intended to cover 500,000 patients market with an optimized design of Synergy for up to 6.0 lpm flow rate. It aims to accommodate the complete spectrum of a patient's circulatory support needs from partial to full. This version has passed the proof-of-concept step and is awaiting animal studies to support the design phase and clinical evaluation. The child support system is also based on modified Synergy for 1.5 to 3 lpm output with a different cannulation. Input cannula will be inserted into the left ventricular apex, while the outflow graft returns blood to the ascending aorta. The infant support system, also called Synergy Nano-Pump, is a super small pump for 0.3 to 1.5 lpm flow and is currently under development for completing proof-of-concept.

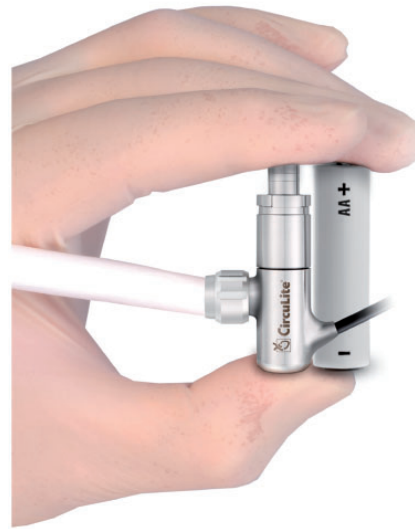

Figure 20. Approximate size of Synergy Micro-Pump (Courtesy of CircuLite, Inc.)

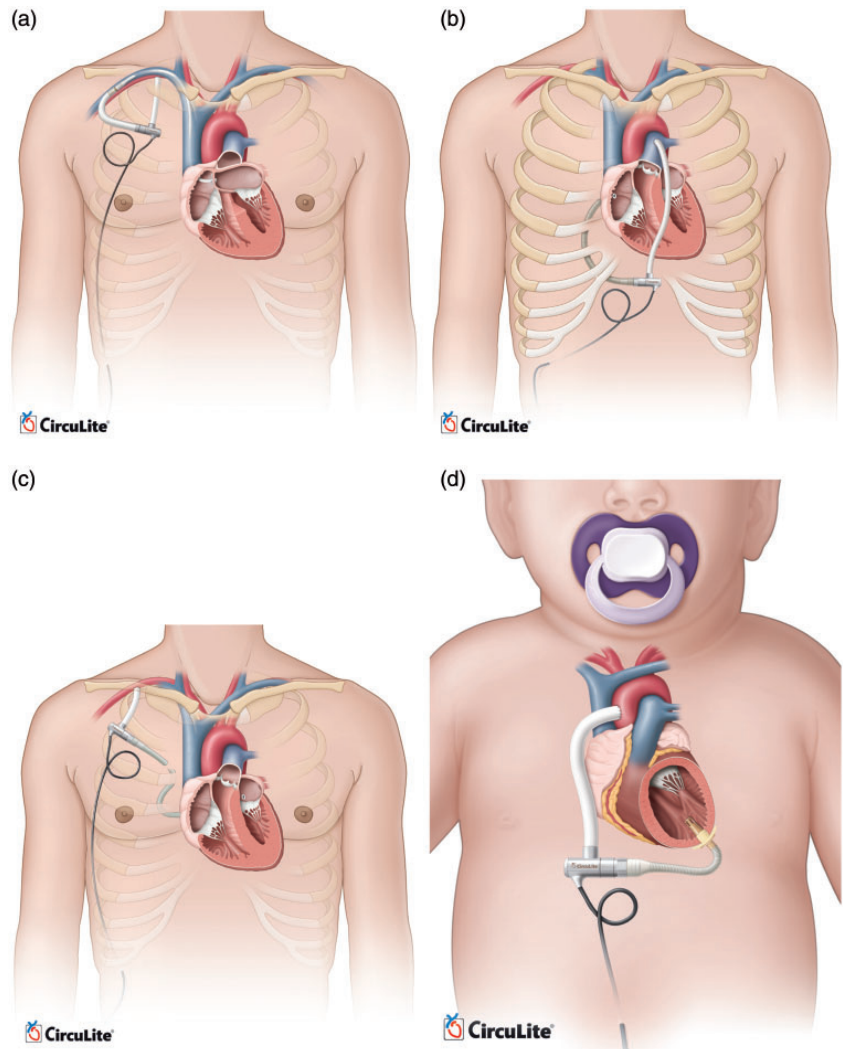

**Figure 21.** Modified versions of CircuLite MicroPump (Courtesy of CircuLite, Inc.). (a) endovascular system, (b) right heart support, (c) all-support and (d) child support.

## Current limitations

Although the MCS field has improved significantly over the past three decades, many challenges are still hindering development of these devices. Stroke remains prevalent.[67] Questions such as optimum time-point for MCS, careful patient selection criteria, and how to maximize the biocompatibility of current systems remain unanswered or under debate. Currently, the decision regarding when to implant an MCS is based on published scientific evidence and INTERMACS values. Ultimately, a decision should be made between MCS implantation at a later time point with the risk of rapidly deteriorating heart failure in the meantime, or earlier MCS with the risk of complications associated with MCS therapy.[68]

The following section provides a summary of the current limitations and challenges, inviting further improvements and developments.

### Device-related issues

Device-related problems such as coagulation disorder, gastrointestinal bleeding, pump thrombosis, device-related infection, and cerebrovascular disorders remain the most critical challenges for MCS systems.[68] Hemorrhage at grafts, thromboembolism at stagnant areas, and hemolysis due to high shearing stress applied to the red blood cells are the main issues with these devices. Heat generation is another matter; if temperature increases high enough, proteins in the blood will denature resulting in irreversible blood damage. Since no valves are employed in these devices, flow back into the heart could worsen the heart failure. The bearing systems for blood pumps should be hemocompatible, reliable, and durable. Mechanical bearings need high flow wash out at the junction between the rotating and stationary parts, which adds to the complexity and difficulty of maintenance. For immersed bearings, the viscous friction in the gap between the rotating impellers and the stationary housing is very high which increases the risk of blood damage. The disadvantage of fully magnetically levitated bearings is that they are highly dependent on the electromagnetic feedback control. If this connection is lost, the rotor crashes in the housing.

### Pulsatility

Patients supported by continuous flow pumps often undergo physiological alterations that are not well understood.[7] Adverse clinical outcomes such as aortic insufficiency, thrombosis, exacerbation of right heart failure, and bleeding remain a significant problem after CF-VAD implantation.[69] Although the long-term effects on survival are unclear,[70] it has been shown that the aortic insufficiency associated with CF-VADs can result in insufficient flow, increased thrombogenicity, and reduced chance of myocardial recovery.[71,72] Pulsation may allow more physiologic unloading of the left ventricle by allowing unloading only during a portion of the cardiac cycle, as opposed to the full cardiac cycle in current CF-VADs, thus reducing pulmonary hypertension and right heart failure.[73] Although some CF-VAD recipients retain some degree of pulsatility because of the improvements in the unloaded heart, others can show no significant pulse, the consequences of which are not clear yet.

### Accurate flow control

Accurate flow control on the current CF-VADs for achieving pulsatility remains a challenge. Speed modulation may offer a potential solution to create pulsatility and prevent these undesired consequences.[45] However, the performance of the flow control algorithm is highly dependent of the accuracy and resolution of the flow measurement. Since direct flow measurement is often impossible, an estimation algorithm is used to predict the pump output based on the pump speed and input power. Accurate, miniaturized, and biocompatible flow feedback loops are lacking. Heartmate III, currently under investigation in USA, has shown promising results for improved flow accuracy and generating physiologic pulsatility.[3]

Using two continuous flow pumps for biventricular support is still facing major challenges. There is no algorithm yet for adjusting the left and right flows by automatically and interdependently adjusting the pump speeds. Therefore, pumps still need two separate controllers. The necessary manual adjustments complicate the care of patients on BiVAD support.[7] The latest versions of CFPs are devices with magnetically levitated and hydrodynamic non-contact impeller suspension systems.[74–76]

### Thrombogenicity

Despite recent advances in MCS development, most notably the CF-VADs, their long-term use remains limited by the thrombotic complications. High shear stress and recirculating flow patterns can lead to platelet activation. Blood contact with artificial surfaces can also stimulate blood clotting. A DT trial on HeartMate II in 2009 showed that 11% of patients suffered a stroke after implantation due to increased shear stress on blood cells.[77] A considerable rate of thromboembolic complications was also reported for the AbioCor TAH.[78] The non-physiologic flow patterns in CF-VADs are considered as one of the major contributors to hemostatic response. Flow patterns along with the artificial surface determine where thrombi will form, its size and composition, and whether it will remain local or it will embolize.[79] Valve designs and hemodynamics also affect the platelet activation.[80] CF-VADs are much smaller than PF-VADs and have smaller gaps between components of the pump. In older generation PF-VADs, any thrombus created in the pump could be dislocated, possibly leading to a stroke. The smaller size of CF-VADs makes them susceptible to thrombosis of the entire pump, where the clot stays in the device and leads to increased hemolysis and device malfunction.[81] Despite aggressive anticoagulant therapy, MCSs can change the coagulation through activating platelets by generating non-physiologic pulsatility, eddy formation, turbulence, or exposure times along platelet trajectories. Development of optimization methods such as DTE has shed more light on identifying thrombogenic regions. This mathematical method combines CFD with particle tracking and in-silico device modeling to determine stress of individual platelets.[79] Platelet activity state (PAS) also facilitates measurement of thrombogenicity of devices.[82] Development and experimental validation of these methods are still an ongoing challenge. Management protocols for MCSs are usually institution dependent, with various factors related to the clinician's experience. Anticoagulation therapy has to be individually tailored. Otherwise, patients may not receive the correct dose of anticoagulant. A significant variation exists between over-anticoagulation and under-anticoagulation in a patient population where coagulation response to the MCS varies considerably between individuals. Over-anticoagulation therapy has often been associated with gastrointestinal bleeding and intracranial hemorrhage. Common complications due to under-anticoagulation include pump thrombosis, hemolysis, and embolic strokes.

### Hemolysis

CF-VADs have demonstrated improved survival and reduced complications compared with PF-VADs. Nevertheless, a study in 2014 reported 18 out of 100 patients receiving HeartMate II were diagnosed with hemolysis.[83] They concluded that hemolysis is associated with very high one-year mortality that is more than two times greater than that observed for the non-hemolyzing patients. Treatments involve MCS exchange or explantation, listing for transplantation, or intensifying anticoagulation therapies. Besides thrombus formation inside the device, a few other hemolysis risk factors are:

- Increased shear stress on red blood cells due to device malpositioning, kinking on the outflow graft/cannula. These cases are often followed by an abrupt change in the LVAD performance such as power spike, flow rate, or pulsatility index
- Dehydration of underfilled left ventricle with increased inlet velocity
- Regurgitant or stenotic valves
- Transfusion-related to immune- and non-immune-mediated hemolysis

Due to the inherent complexity of patients' responses to anticoagulant therapies, clinicians may have to choose between thrombosis, bleeding, or sometimes both.[81] Therefore, achieving the optimal anticoagulation therapy remains an ongoing challenge. Unfortunately, researchers have limited access to data regarding the prognosis of patients with hemolysis. More single-center studies on the incident and clinical outcomes associated with thrombosis, hemolysis, gastrointestinal bleeding, and infection have been reported over the last 5 to 10 years.[81,83–87]

### Gastrointestinal bleeding

Multiple reports have been published on the gastrointestinal (GI) bleeding in patients who receive MCS.[88–92] In a narrative review in 2013, 20.5% of LVAD recipients out of 1543 patients developed GI bleeding.[90] This complication has been more pronounced in CF-VADs compared with PF-VADs. Bleeding can occur throughout the GI tract. Although the bleeding is usually manageable, the definite bleeding site remains unidentified. Lack of enough knowledge about the bleeding site hinders the MCS technology refinements and improvements. Despite some GI bleedings are directly related to the device itself, LVAD replacement is seldom necessary.[93] More research is needed on the etiology of GI bleeding as well as the proper anticoagulation management before and after GI bleeding. The role of anticoagulation acquired von Willebrand syndrome and platelet dysfunction due to altered hemodynamics are not clearly understood yet.

### Infection

Infection occurs in up to 60% of MCS recipients.[94,95] MCS-related infections, bacteremia, and fungemia are common in these patients. Major MCS-related infections are percutaneous driveline, pump pocket, and cannula. The percutaneous driveline, particularly at the exit site, is quite susceptible to infection. Visual inspection of driveline for diagnosing infections may be hindered by purulent discharge and surrounding cellulitis. It is reported that even with infection, systemic signs such as fever, elevated inflammatory markers, or leukocytosis may not be present.[96] Treatment of infection may require removal of the device, since the entire device can be susceptible to biofilm formation.[97] Cannula and pump infections are among the most serious MCS-related infections. Infection along the cannula or inside the pump can lead to pump failure due to flow blockage. Besides the differences in the MCS technologies themselves, heart failure risk score, older age, diabetes, renal failure, and nutritional status are known as the additional factors associated with the overall infection.

CF-VADs have shown less MCS-related infections compared with PF-VADs due to smaller pump size and driveline caliber. However, sepsis and non-MCS-related infections still cause the majority of infections during the first 90 days of implantation. MCS-related infections, especially driveline and pump pocket infections, comprise the majority of infections after 90 days of implantation.[98,99] Although infection complications of MCS therapy have been described for many years, current recommendations for monitoring infection after MCS implantation are based on observational studies and expert opinions. Randomized controlled trials for studying the MCS-related and non-MCS-related infections after implantation seem essential. An ongoing effort towards reducing MCS-related infections is realization of TETS. TETS can recharge the batteries wirelessly, so patients are neither tethered to a large air pumping console nor pierced by drivelines and tubes.[100–102]

## Declaration of conflicting interests
The author(s) declared no potential conflicts of interest with respect to the research, authorship, and/or publication of this article.

## Funding
The author(s) would like to acknowledge the funding of Ohio Third Frontier Grant WP 10-010 for this project.

## Guarantor
MH is the guarantor of this paper.

## Contributorship
MH designed the work, performed the literature review, and drafted the article. ME revised it technically as an engineering professional. RG and MB revised the manuscript critically from a medical perspective. All authors contributed to and approved the final version.

## References

1. Garbade J, Barten MJ, Bittner HB, et al. Heart transplantation and left ventricular assist device therapy: two comparable options in end-stage heart failure? *Clin Cardiol* 2013; 36: 378–82.
2. Agarwal S and High KM. Newer-generation ventricular assist devices. *Best Pract Res Clin Anaesthesiol* 2012; 26: 117–130.
3. Schumer EM, Black MC, Monreal G, et al. Left ventricular assist devices: current controversies and future directions. *Eur Heart J* 2016; 37: 3434–3439.
4. Loforte A, Musumeci F, Montalto A, et al. Use of mechanical circulatory support devices in end-stage heart failure patients. *J Card Surg* 2014; 29: 717–722.
5. Slaughter MS, Bostic R, Tong K, et al. Temporal changes in hospital costs for left ventricular assist device implantation. *J Card Surg* 2011; 26: 535–541.
6. Dang N, Topkara V, Mercando M, et al. Right heart failure after left ventricular assist device implantation in patients with chronic congestive heart failure. *J Heart Lung Transplant* 2006; 25: 1–6.
7. Mallidi HR, Anand J and Cohn WE. State of the art of mechanical circulatory support. *Tex Heart Inst J* 2014; 41: 115–20.
8. Araki K, Anai H, Oshikawa M, et al. Invitro performance of a centrifugal, a mixed Flow, and an axial flow blood pump. *Artif Organs* 1998; 22: 366–370.
9. Fraser KH, Zhang T, Taskin ME, et al. A Quantitative comparison of mechanical blood damage parameters in rotary ventricular assist devices: shear stress, exposure time and hemolysis index. *J Biomech Eng* 2012; 134: 081002.
10. Antaki JF, Ricci MR, Verkaik JE, et al. PediaFlow Maglev ventricular assist device: a prescriptive design approach. *Cardiovasc Eng Technol* 2010; 1: 104–121.
11. Pagani FD. Continuous-flow rotary left ventricular assist devices with "3rd Generation" design. *Semin Thorac Cardiovasc Surg* 2008; 20: 255–263.
12. Moazami N, Fukamachi K, Kobayashi M, et al. Axial and centrifugal continuous-flow rotary pumps: a translation from pump mechanics to clinical practice. *J Heart Lung Transplant* 2013; 32: 1–11.
13. Reul HM and Akdis M. Blood pumps for circulatory support. *Perfusion* 2000; 15: 295–311.
14. Frazier OH, Khalil HA, Benkowski RJ, et al. Optimization of axial-pump pressure sensitivity for a continuous-flow total artificial heart. *J Heart Lung Transplant* 2010; 29: 687–691.
15. LaRose JA, Tamez D, Ashenuga M, et al. Design concepts and principle of operation of the HeartWare ventricular assist system. *ASAIO J* 2010; 56: 285–289.
16. Wu ZJ, Gottlieb RK, Burgreen GW, et al. Investigation of fluid dynamics within a miniature mixed flow blood pump. *Stand* 2001; 31: 615–629.
17. Carrier M, Garon A, Camarero R, et al. Dual inlet mixed-flow blood pump. US Patent 20050107657 A1, USA, 2005.
18. Ising M, Warren S, Sobieski MA, et al. Flow modulation algorithms for continuous flow left ventricular assist devices to increase vascular pulsatility: a computer simulation study. *Cardiovasc Eng Technol* 2011; 2: 90–100.
19. Sundareswaran KS, Reichenbach SH, Masterson KB, et al. Low bearing wear in explanted HeartMate II left ventricular assist devices after chronic clinical support. *Asaio J* 2013; 59: 41–45.
20. Carriker JW and Wampler RK. Percutaneous axial flow blood pump. US Patent 4944722, USA, 1990.
21. Benkowski RJ and Hudson L. Rotary blood pump. US Patent US 200910143635 A1, USA, 2009.
22. Moise JC, Wampler RK and Butler KC. Chronic Ventricular Assist System. US Patent 4908012, USA, 1990.
23. Moise JC. Magnetically suspended rotor axial now blood pump. US Patent 4779614, USA, 1988.
24. Olsen DB, Bramm G and Novak P. Magnetically suspended and rotated impeller pump apparatus and method. US Patent 4688998, USA, 1987.
25. Bramm G and Novak P. Magnetic rotor bearing. US Patent 4763032, USA, 1988.
26. Weber N, Wendel HP and Ziemer G. Hemocompatibility of heparin-coated surfaces and the role of selective plasma protein adsorption. *Biomaterials* 2002; 23: 429–39.
27. Okkema AZ, Yu XH and Cooper SL. Physical and blood contacting characteristics of propyl sulphonate grafted biomer. *Biomaterials* 1991; 12: 3–12.
28. Han DK, Lee NY, Park KD, et al. Heparin-like anticoagulant activity of sulphonated poly(ethylene oxide) and sulphonated poly(ethylene oxide)-grafted polyurethane. *Biomaterials* 1995; 16: 467–471.
29. Sandhu S and Luthra A. *Developments in Biointeracting Materials for Medical Application. Business Briefing: Medical Device Manufacturing and Technology*. Technical Report. London, UK: Touch Briefings, 2004, pp.1–4.
30. Timms D. A review of clinical ventricular assist devices. *Med Eng Phys* 2011; 33: 1041–1047.

31. Bozeman JR, Akkerman JW, Aber GS, et al. Rotary blood pump. US Patent 5527159, USA, 1996.
32. Benkowski RJ, Kiris C, Kwak D, et al. Rotary blood pump. US Patent 5947892, USA, 1999.
33. Goldstein D, Zucker M, Arroyo L, et al. The Micromed-Debakey VAD: initial American experience. *J Heart Lung Transplant* 2003; 22: S218.
34. Chiu WC, Girdhar G, Xenos M, et al. Thromboresistance comparison of the HeartMate II ventricular assist device with the device thrombogenicity emulation – optimized HeartAssist 5 VAD. *J Biomech Eng* 2014; 136: 021014.
35. Burke DJ, Burke E, Parsaie F, et al. The Heartmate II: design and development of a fully sealed axial flow left ventricular assist system. *Artif Organs* 2001; 25: 380–385.
36. John R. Current axial-flow devices – the HeartMate II and Jarvik 2000 left ventricular assist devices. *Semin Thorac Cardiovasc Surg* 2008; 20: 264–272.
37. Thoratec Corporation. HeartMate II LVAS instructions for use, featuring GoGear system components (105747.A). Technical report, Thoratec Corporation, Pleasanton, CA, USA, 2010.
38. Hoshi H, Shinshi T and Takatani S. Third-generation blood pumps with mechanical noncontact magnetic bearings. *Artif Organs* 2006; 30: 324–338.
39. Huber CH, Tozzi P, Hurni M, et al. No drive line, no seal, no bearing and no wear: magnetics for impeller suspension and flow assessment in a new VAD. *Interact Cardiovasc Thorac Surg* 2004; 3: 336–340.
40. Schmid C, Tjan TDT, Etz C, et al. First clinical experience with the Incor left ventricular assist device. *J Heart Lung Transplant* 2005; 24: 1188–1194.
41. Frazier OH and Jacob LP. Small pumps for ventricular assistance: progress in mechanical circulatory support. *Cardiol Clin* 2007; 25: 553–564; vi.
42. Weber DM, Raess DH, Henriques JP, et al. Principles of Impella cardiac support. *Princ Hemodynamics* 2009; (September): 3–16.
43. Seyfarth M, Sibbing D, Bauer I, et al. A randomized clinical trial to evaluate the safety and efficacy of a percutaneous left ventricular assist device versus intra-aortic balloon pumping for treatment of cardiogenic shock caused by myocardial infarction. *J Am Coll Cardiol* 2008; 52: 1584–1588.
44. Siess T, Nix C and Boensch S. *Method for calibrating a pressure sensor or a flow sensor at a rotary pump*. US Patent 7010954 B2, 2006.
45. Dixon SR, Henriques JP, Mauri L, et al. A prospective feasibility trial investigating the use of the Impella 2.5 system in patients undergoing high-risk percutaneous coronary intervention (The PROTECT I Trial). *JACC Cardiovasc Interv* 2009; 2: 91–96.
46. Hetzer R, Krabatsch T, Stepanenko A, et al. Long-term biventricular support with the heartware implantable continuous flow pump. *J Heart Lung Transplant* 2010; 29: 822–824.
47. Strueber M, Meyer AL, Malehsa D, et al. Successful use of the HeartWare HVAD rotary blood pump for biventricular support. *J Thorac Cardiovasc Surg* 2010; 140: 936–937.
48. Slaughter M, Sobieski M and II D. HeartWare miniature axial-flow ventricular assist device: design and initial feasibility test. *Tex Heart Inst J* 2009; 36: 12–16.
49. Aaronson KD, Slaughter MS, Miller LW, et al. Use of an intrapericardial, continuous-flow, centrifugal pump in patients awaiting heart transplantation. *Circulation* 2012; 125: 3191–3200.
50. Bourque K, Gernes DB, Loree HM 2nd, et al. HeartMate III: pump design for a centrifugal LVAD with a magnetically levitated rotor. *Asaio J* 2001; 47: 401–405.
51. Shah KB, Tang DG, Cooke RH, et al. Implantable mechanical circulatory support: demystifying patients with ventricular assist devices and artificial hearts. *Clin Cardiol* 2011; 34: 147–152.
52. Nguyen DQ and Thourani VH. Third-generation continuous flow left ventricular assist devices. *Innovations (Phila)* 2010; 5: 250–258.
53. Farrar DJ, Bourque K, Dague CP, et al. Design features, developmental status, and experimental results with the Heartmate III centrifugal left ventricular assist system with a magnetically levitated rotor. *ASAIO J* 2007; 53: 310–315.
54. Thoratec Corporation. VAD trials and outcomes – clinical outcomes – Thoratec PVAD, www.thoratec.com (2015, accessed 2 August 2017).
55. Bermudez C, Minakata K and Kormos RL. *Redo cardiac surgery in adults*. New York, NY: Springer, 2012.
56. Mueller JP, Kuenzli A, Reuthebuch O, et al. The CentriMag: a new optimized centrifugal blood pump with levitating impeller. *Heart Surg Forum* 2004; 7: E477–E480.
57. De Robertis F, Birks EJ, Rogers P, et al. Clinical performance with the Levitronix Centrimag short-term ventricular assist device. *J Heart Lung Transplant* 2006; 25: 181–186.
58. Westaby S, Balacumaraswami L, Evans BJ, et al. Elective transfer from cardiopulmonary bypass to centrifugal blood pump support in very high-risk cardiac surgery. *J Thorac Cardiovasc Surg* 2007; 133: 577–578.
59. Hijikata W, Sobajima H, Shinshi T, et al. Disposable maglev centrifugal blood pump utilizing a cone-shaped impeller. *Artif Organs* 2010; 34: 669–677.
60. Hijikata W, Shinshi T, Asama J, et al. A magnetically levitated centrifugal blood pump with a simple-structured disposable pump head. *Artif Organs* 2008; 32: 531–540.
61. Yamazaki K, Kihara S, Akimoto T, et al. EVAHEART: an implantable centrifugal blood pump for long-term circulatory support. *Jpn J Thorac Cardiovasc Surg* 2002; 50: 461–465.
62. Snyder T, Tsukui H, et al. Preclinical biocompatibility assessment of the EVAHEART ventricular assist device: coating comparison and platelet activation. *J Biomed Mater Res Part A* 2007; 81: 85–92.
63. Yamazaki K, Saito S, Kihara S, et al. Completely pulsatile high flow circulatory support with a constant-speed centrifugal blood pump: mechanisms and early clinical observations. *Gen Thorac Cardiovasc Surg* 2007; 55: 158–162.

64. Yamazaki K, Saito S, Nishinaka T, et al. 517: Japanese clinical trial results of an implantable centrifugal blood pump EVAHEART. *J Heart Lung Transplant* 2008; 27: S246.
65. McCarthy PM. Partial mechanical cardiac support: part of the solution or part of the problem? *J Am Coll Cardiol* 2009; 54: 87–88.
66. Spiliopoulos K, Giamouzis G, Karayannis G, et al. Current status of mechanical circulatory support: a systematic review. *Cardiol Res Pract* 2012; 2012: 1–12.
67. Kirklin JK, Naftel DC, Kormos RL, et al. Fifth INTERMACS annual report: risk factor analysis from more than 6,000 mechanical circulatory support patients. *J Heart Lung Transplant* 2013; 32: 141–156.
68. Puehler T, Ensminger S, Schoenbrodt M, et al. Mechanical circulatory support devices as destination therapy – current evidence. *Ann Cardiothorac Surg* 2014; 3: 513–524.
69. Frazier OH. Unforeseen consequences of therapy with continuous-flow pumps. *Circ Heart Fail* 2010; 3: 647–649.
70. Hiraoka A, Cohen JE, Shudo Y, et al. Evaluation of late aortic insufficiency with continuous flow left ventricular assist device. *Eur J Cardio-Thoracic Surg* 2015; 48: 400–406.
71. Imamura T, Kinugawa K, Nitta D, et al. Advantage of pulsatility in left ventricular reverse remodeling and aortic insufficiency prevention during left ventricular assist device treatment. *Circ J* 2015; 79: 1994–1999.
72. Cowger J, Rao V, Massey T, et al. Comprehensive review and suggested strategies for the detection and management of aortic insufficiency in patients with a continuous-flow left ventricular assist device. *J Heart Lung Transplant* 2015; 34: 149–157.
73. John R, Lee S, Eckman P, et al. Right ventricular failure: a continuing problem in patients with left ventricular assist device support. *J Cardiovasc Transl Res* 2010; 3: 604–611.
74. DeBakey ME. A miniature implantable axial flow ventricular assist device. *Ann Thorac Surg* 1999; 68: 637–640.
75. Potapov EV, Loebe M, Nasseri BA, et al. Pulsatile Flow in Patients With a Novel Nonpulsatile Implantable Ventricular Assist Device. *Circulation* 2000; 102(Suppl 3): III-183–III-187.
76. Timms D. *Design, development and evaluation of centrifugal ventricular assist devices*. PhD Thesis, School of Mechanical, Manufacturing and Medical Engineering, Queensland University of Technology, 2005.
77. Slaughter MS, Rogers JG, Milano CA, et al. Advanced heart failure treated with continuous-flow left ventricular assist device. *N Engl J Med* 2009; 361: 2241–2251.
78. Frazier OH, Dowling RD, Gray LA, et al. The total artificial heart: where we stand. *Cardiology* 2004; 101: 117–121.
79. Bluestein D, Einav S and Slepian MJ. Device thrombogenicity emulation: a novel methodology for optimizing the thromboresistance of cardiovascular devices. *J Biomech* 2013; 46: 338–344.
80. Laas J, Kseibi S, Perthel M, et al. Impact of high intensity transient signals on the choice of mechanical aortic valve substitutes. *Eur J Cardiothorac Surg* 2003; 23: 93–96.
81. Tchantchaleishvili V, Sagebin F, Ross RE, et al. Evaluation and treatment of pump thrombosis and hemolysis. *Ann Cardiothorac Surg* 2014; 3: 490–495.
82. Jesty J, Yin W, Perrotta P, et al. Platelet activation in a circulating flow loop: combined effects of shear stress and exposure time. *Platelets* 2003; 14: 143–149.
83. Ravichandran AK, Parker J, Novak E, et al. Hemolysis in left ventricular assist device: a retrospective analysis of outcomes. *J Heart Lung Transplant* 2014; 33: 44–50.
84. Whitson Ba, Eckman P, Kamdar F, et al. Hemolysis, pump thrombus, and neurologic events in continuous-flow left ventricular assist device recipients. *Ann Thorac Surg* 2014; 97: 2097–2103.
85. Heilmann C, Geisen U, Benk C, et al. Hemolysis in patients with ventricular assist devices (VAD): systems differ significantly. *Thorac Cardiovasc Surg* 2009; 56(S 01): V152.
86. Hasin T, Deo S, Maleszewski JJ, et al. The role of medical management for acute intravascular hemolysis in patients supported on axial flow LVAD. *ASAIO J* 2014; 60: 9–14.
87. Cowger JA, Romano MA, Shah P, et al. Hemolysis: a harbinger of adverse outcome after left ventricular assist device implant. *J Heart Lung Transplant* 2014; 33: 35–43.
88. Islam S, Cevik C, Madonna R, et al. Left ventricular assist devices and gastrointestinal bleeding: a narrative review of case reports and case series. *Clin Cardiol* 2013; 36: 190–200.
89. Harvey L, Holley CT and John R. Gastrointestinal bleed after left ventricular assist device implantation: incidence, management, and prevention. *Ann Cardiothorac Surg* 2014; 3: 475–479.
90. Suarez J, Patel CB, Felker GM, et al. Mechanisms of bleeding and approach to patients with axial-flow left ventricular assist devices. *Circ Heart Fail* 2011; 4: 779–784.
91. Eckman PM and John R. Bleeding and thrombosis in patients with continuous-flow ventricular assist devices. *Circulation* 2012; 125: 3038–3047.
92. Crow S, John R, Boyle A, et al. Gastrointestinal bleeding rates in recipients of nonpulsatile and pulsatile left ventricular assist devices. *J Thorac Cardiovasc Surg* 2009; 137: 208–215.
93. Miller ED, Steidley DE, Arabia FA, et al. Gastric erosion associated with left ventricular assist device: new technology, new complication. *Gastrointest Endosc* 2009; 70: 181–183.
94. Califano S, Pagani FD and Malani PN. Left ventricular assist device-associated infections. *Infect Dis Clin North Am* 2012; 26: 77–87.
95. Gordon RJ, Quagliarello B and Lowy FD. Ventricular assist device-related infections. *Lancet Infect Dis* 2006; 6: 426–437.
96. Rose E, Gelijns A, Moskowitz A, et al. Long-term use of a left ventricular assist device for end-stage heart failure. *N Engl J Med* 2001; 345: 1435–1443.

97. Angud M. Left ventricular assist device driveline infections. *AACN Adv Crit Care* 2015; 26: 300–305.
98. Topkara VK, Kondareddy S, Malik F, et al. Infectious complications in patients with left ventricular assist device: etiology and outcomes in the continuous-flow era. *Ann Thorac Surg* 2010; 90: 1270–1277.
99. Koval CE and Rakita R. Ventricular assist device related infections and solid organ transplantation. *Am J Transplant* 2013; 13(Suppl 4): 348–354.
100. Zarinetchi F, Hart RM, Verga MG, et al. Transcutaneous energy transfer device with magnetic field protected components in secondary coil. US Patent 6496733 B2, USA, 2002.
101. Miller JA. Transcutaneous energy transfer device. US Patent 5350413, USA, 1994.
102. Dissanayake T, Budgett D, Hu AP, et al. Transcutaneous Energy Transfer System for Powering Implantable Biomedical Devices. In: Chwee Teck Lim and James CH Goh (eds) *13th International Conference on Biomedical Engineering, Part of the IFMBE Proceedings book series (volume 23)*, Singapore, 3–6 December 2008, pp.235–239.